\documentclass[%
reprint,
amsmath,amssymb, 
aps,
prd,
]{revtex4-1}

\usepackage{verbatim}
\usepackage{graphics}
\usepackage{graphicx}
\usepackage{epsfig}
\usepackage{subfigure}
\usepackage{dcolumn}
\usepackage{bm}
\usepackage{amsmath}
\usepackage{mathrsfs}
\usepackage[retainorgcmds]{IEEEtrantools}

\newcommand{\mnras}{MNRAS}
\newcommand{\apjl}{ApJ}
\newcommand{\apjs}{ApJS}
\newcommand{\aj}{AJ}
\newcommand{\aap}{A\&A}
\newcommand{\jcap}{JCAP}
\newcommand{\na}{New Astronomy}
\newcommand{\physrep}{Physics Reports}

\newcommand{\ud}{\mathrm{d}}

\begin{document}


\title{Cosmological Constraints on Bose-Einstein-Condensed Scalar Field Dark Matter}

\author{Bohua Li}
 \email{bohuali@astro.as.utexas.edu}
\author{Tanja Rindler-Daller}
\thanks{present address: Department of Physics, University of Michigan, 
450 Church Street, Ann Arbor, MI 48109, USA. daller@umich.edu}
\author{Paul R. Shapiro}%
 \email{shapiro@astro.as.utexas.edu}
\affiliation{%
 Department of Astronomy and Texas Cosmology Center, The University
 of Texas at Austin, 2515 Speedway C1400, Austin, TX 78712, USA
}%

\date{\today}

\begin{abstract}

Despite the great successes of the Cold Dark Matter (CDM) model in
explaining a wide range of observations of the global evolution and
the formation of galaxies and large-scale structure in the Universe,
the origin and microscopic nature of dark matter is still unknown.
The most common form of CDM considered to-date is that of Weakly
Interacting Massive Particles (WIMPs), but, so far, attempts to
detect WIMPs directly or indirectly have not yet succeeded, and the
allowed range of particle parameters has been significantly
restricted. Some of the cosmological predictions for this kind of
CDM are even in apparent conflict with observations 
(e.g. cuspy-cored halos and an overabundance of satellite dwarf galaxies). 
For these reasons, it is important to consider the consequences of
different forms of CDM. We focus here on the hypothesis that the
dark matter is comprised, instead, of ultralight bosons that form a
Bose-Einstein Condensate, described by a complex scalar field, 
for which particle number per unit comoving volume is conserved.
We start from the Klein-Gordon and Einstein field equations to
describe the evolution of the Friedmann-Robertson-Walker 
universe in the presence of this kind of dark matter. We find that,
in addition to the radiation-, matter-, and $\Lambda$-dominated 
phases familiar from the standard CDM model, there
is an earlier phase of scalar-field-domination, which is special to
this model. In addition, while WIMP CDM is non-relativistic at all
times after it decouples, the equation of state of Bose-Einistein condensed 
scalar field dark matter (SFDM) is found to be relativistic at early times,
evolving from stiff ($\bar p = \bar\rho$) to radiationlike ($\bar p =
\bar \rho/3$), before it becomes non-relativistic and CDM-like at late
times ($\bar p = 0$). The timing of the transitions between these phases
and regimes is shown to yield fundamental constraints on the SFDM
model parameters, particle mass $m$ and self-interaction coupling
strength $\lambda$. We show that SFDM is compatible with observations of the
evolving background universe, by deriving the range of particle
parameters required to match observations of the cosmic microwave
background (CMB) and the abundances of the light elements produced
by big bang nucleosynthesis (BBN), including $N_{\rm{eff}}$, 
the effective number of neutrino species, and the epoch of
matter-radiation equality $z_{\rm{eq}}$. This yields $m\ge2.4\times10^{-21}$eV$/c^2$ and 
$9.5\times10^{-19}\rm{eV^{-1}cm^3}\le\lambda/(mc^2)^2\le4\times10^{-17}\rm{eV^{-1}cm^3}$. 
Indeed, our model can accommodate current observations 
in which $N_{\rm{eff}}$ is higher at the BBN epoch than at $z_{\rm{eq}}$, probed by the CMB,
which is otherwise unexplained by the standard CDM model involving WIMPs.
We also show that SFDM without self-interaction (also called ``fuzzy dark matter'') is not able to 
comply with the current constraints from BBN within 68\% confidence, and is therefore disfavored.

\begin{description}
\item[PACS numbers]
98.80.-k, 95.35.+d, 98.80.Cq, 98.80.Ft

\end{description}
\end{abstract}

\maketitle


\section{Introduction} \label{sec:intro}

\subsection{Cold dark matter}

Since the discovery of the accelerating expanding Universe,
$\Lambda\text{CDM}$ has become the standard cosmological model as
supported by various astronomical observations. Cosmic microwave background (CMB) observations
have shown that about 25\% of the energy density of the present
Universe is comprised of non-baryonic cold dark matter. Cold dark
matter (CDM) does not interact under electromagnetism and the strong
force, and moves non-relativistically, thus acting like cold,
pressureless dust in the present Universe. Despite these
characteristics, its particle nature is still unknown and no
candidate can be found within the Standard Model of particle physics
(SM). So far, diverse extensions of the SM have predicted candidate
particles for CDM, among which the most popular ones at present are
in the form of weakly interacting massive particles (WIMPs) (see Refs. \cite{GW1985, ST1986, DFS1986}). 
WIMPs are collisionless and massive ($>$ GeV). 

The standard collisionless CDM, in a universe perturbed by Gaussian-random-noise 
primordial density fluctuations with a nearly scale-independent
primordial power spectrum, provides a well-accepted scenario for
cosmic structure formation: the hierarchical clustering of dark matter
fluctuations and the infall of baryons into CDM potential wells
after recombination, to form galaxies. Despite the fact that this
story line is in good agreement with many observational constraints,
including CMB anisotropy \cite{EWMAP7, WMAP9, Planck}, large-scale
structure \cite{BAOBOSS9, 2PCFBOSS9} and the general properties of 
dark-matter-dominated halos \cite{Walker2009, Newman2013, Vikhlinin2009}, 
some crucial issues on small scales 
are subject to controversy (see Ref. \cite{WD2013CDMP} for a recent, brief review). 
First, hierarchical clustering in the standard CDM model overpredicts 
the number of substructures in a halo the size of the Local Group by an order of magnitude 
as compared with the number of satellite galaxies observed in the Local Group, 
a discrepancy referred to as the ``missing satellite problem'' 
(see Refs. \cite{Klypin1999, Moore2, Aquarius2008, MSII2009}). 
Second, the density profiles of collisionless CDM halos in N-body simulations 
show a universal profile with a central cusp ($\sim r^{-1}$ in the NFW profile \cite{NFW}), 
while observations of low-surface brightness galaxies and dwarf galaxies 
mostly favor a flat central slope. This has been known as the ``cuspy core problem'' 
(see Refs. \cite{Moore1, dB2001, Oh2008, Amorisco2012}). 
Furthermore, current dark matter detection experiments, both direct and indirect ones, 
have not yet discovered any compelling signals of WIMPs \cite{Bauer2013DMDETECT}. 
As a matter of fact, while WIMPs are mostly expected to be 
the lightest supersymmetric particle 
in the Minimal Supersymmetric Standard Model (MSSM), e.g., neutralinos \cite{GK2000}, 
recent data from the Large Hadron Collider has found no evidence 
of a deviation from the SM on GeV scales, 
significantly restricting the allowed region of MSSM parameters \cite{ATLAS2013, CMS2013}. 
All these facts taken together, it is evident that the microscopic nature of dark matter 
is sufficiently unsettled as to justify the consideration of alternative candidates 
for the CDM paradigm, especially in the hope of resolving the above difficulties.

\subsection{Bose-Einstein-condensed ultra-light particles as dark matter candidate} \label{sec:BEC}

We assume that the dark matter particles are described by a spin-0
scalar field (`scalar field dark matter', for short; henceforth, SFDM) with a possible self-interaction. 
In fact, one type of bosonic particle suggested as a major candidate for dark matter 
is the QCD axion. It is the pseudo-Nambu-Goldstone boson in
the Peccei-Quinn mechanism, proposed as a dynamical solution to the
strong CP-problem in QCD. For the axion to be CDM, it has to be very
light, $m\sim10^{-5}$ eV$/c^2$ \cite{Weinberg, Sikivie2012}.

In addition to the QCD axion, several fundamental scalar fields have
been predicted by a variety of unification theories, e.g., string
theories and other multi-dimensional theories \cite{Carroll1998, ADD1999, Kallosh2002, Arvanitaki2010}. 
The bosonic particles envisaged are typically ultralight, with masses down to the order of $10^{-33}$ eV$/c^2$. 
This suggests an ultrahigh phase-space density,
leading to the possibility of formation of a Bose-Einstein
condensate (BEC), i.e., a macroscopic occupancy of the many-body
ground state. In principle, for a fixed number of (locally) thermalized
identical bosons, a BEC will form if $n\lambda^3_{\rm{deB}}\gg 1$, where
$n$ is the number density and $\lambda_{\rm{deB}}$ is the
de Broglie wavelength. This is equivalent to there being a critical temperature
$T_c$, below which a BEC can form. 

For a non-relativistic, ideal (i.e. non-interacting) boson gas, the well-known result for $T_c$ is  
\begin{equation}\label{equation:nibg}
	T_c=\frac{2\pi \hbar^2}{mk_B}\left(\frac{n}{\zeta(3/2)}\right)^{2/3}, 
\end{equation}
which was used, for example, by Refs. \cite{Harko2011, FM2009}. 
Equation (\ref{equation:nibg}) is not an adequate description of the case considered here, however. 
For the ultralight particles with which we are concerned, $k_BT_c/mc^2\gg1$, 
so a fully relativistic treatment is required. 

We are interested in a complex scalar field, 
for which the presence of dark matter results from the asymmetry associated with 
the difference between the number density of bosons and that of their anti-particles, 
a conserved \textit{charge} density in the comoving frame 
(see also Appendix \ref{app:charge} for more discussion about the charge). 
A fully relativistic treatment of Bose-Einstein (`BE') condensation was given 
by Refs. \cite{Kapusta1981} and \cite{HW1982}, 
including the relationship between BE condensation and symmetry breaking of a scalar field. 
Those authors showed that, for an ultra-relativistic ideal \textit{charged} boson gas, 
described by a \textit{complex} scalar field, 
\begin{equation}
	T_c=\frac{(\hbar^3c)^{1/2}}{k_B}\left(\frac{3q}{m}\right)^{1/2},
\end{equation}
where $q$ is the charge per unit \textit{proper} volume. 
This does not, however, take self-interaction into account. 
Reference \cite{HW1982} showed that, in the case of an adiabatically expanding boson gas, 
relevant to cosmology, if the scalar field has a generic quartic self-interaction, 
then the bosons must either be condensed at all temperatures 
(i.e. at all times) or else never form a BEC. In this case, 
the charge per unit \textit{comoving} volume, $Q~(Q=qa^3)$, 
and entropy per unit comoving volume, $S$, are both conserved. 
According to equation (4.7) of that paper, 
a (local) BEC will exist from the beginning and remain at all times, if 
\begin{equation}\label{equation:BECcond}
	\frac{Q}{S}\gg\frac{5}{4\pi^2k_B}\left(\frac{\hat\lambda}{4}\right)^{1/2}, 
\end{equation}
where $\hat\lambda$ is the \textit{dimensionless} coupling strength of the quartic self-interaction, in natural units. 
Our SFDM has essentially zero entropy per unit comoving volume. 
Also, for the small boson masses that we will be considering, 
the conserved charge density in the comoving frame, $Q$, is extremely high, 
given the observed present-day dark matter energy density 
$\bar\rho_{\rm{dm}}(t_0)$, for $Q\approx \bar\rho_{\rm{dm}}(t_0)/(mc^2)$. 
Therefore, we are always in the regime described by inequality (\ref{equation:BECcond}), 
and thus the bosons are fully condensed from the time they are born, i.e., almost all of the bosons
occupy the lowest available energy state. 

Hence, the cosmological Bose-Einstein-condensed SFDM can be described 
by a single (coherent) classical scalar field, of which the value at each point in space
equals to that of the local order parameter \cite{Pitaevskii}. 
Even though the condensation requires Bose-Einstein statistics in
the first place, i.e., local thermalization (see Refs. \cite{SY2009, Erken2012}), 
we argue that thermal decoupling within the bosonic dark matter
can occur when the expansion rate exceeds its thermalization
rate, without disturbing the condensate. Most of the bosons
will stay in the ground state (BEC), and the classical field
(SFDM) remains a good description, analogous to the fact that CMB
photons after decoupling still follow a black-body distribution. 
In summary, we consider the Bose-Einstein condensate 
as an initial condition for our model, such that we can use and trust the effective field description 
throughout the evolution of the universe up to very early times. 

A scalar field description of BEC dark matter has been
studied by several authors before; see, for instance, 
Refs. \cite{Sin, LeeKoh, Goodman, HBG, Harko2011, Magana2012, RS1, Chavanis2012}. 
With regard to the aforementioned initial condition, one may also envisage a scenario 
in which the coherent scalar field is created gravitationally at the end of inflation, 
as has been considered, e.g., by Refs. \cite{Ford, PV1999Q, Peebles2000}. 
On the other hand, it might also be that SFDM was just another scalar field, 
in place along with the inflaton before and during inflation \cite{FHW1992, ALS2002}, 
emanating from yet earlier initial conditions. 
Speculations of that kind are beyond the scope of this paper. 
However, we find some interesting early-time features which will deserve more discussion in due course.

A prime motivation for studying SFDM has been its ability to suppress 
small-scale clustering and hence potentially resolve the dark matter problems mentioned above. 
For non-self-interacting particles, $\lambda_{\rm{deB}} = h/(mv)$ 
sets a natural lower limit to the scale on which equilibrium halos can form, 
where $v$ corresponds to the virial velocity of the galactic halos. 
While this paper shall deal only with the consequences of SFDM for the homogeneous 
background universe, this argument would suggest that there is a lower limit to 
the particle mass for SFDM of $m\gtrsim10^{-22}$ eV$/c^2$, 
since then $\lambda_{\rm{deB}} \lesssim 1$ kpc \cite{RS1, RS2}, 
the core size of the dark matter halo of a typical dwarf spheroidal galaxy 
in the present Universe \cite{Amorisco2013}. If self-interaction of SFDM is included, 
the associated characteristic gravitational equilibrium scale $l_{\rm{SI}}$ is proportional to
$\sqrt{\lambda/(mc^2)^2}$, where $\lambda$ is the \textit{dimensional} coupling strength of the quartic self-interaction 
(related to $\hat\lambda$ by $\lambda\equiv\hat\lambda\frac{\hbar^3}{m^2c})$, 
i.e., $l_{\rm{SI}} \simeq 1$ kpc if $\lambda/(mc^2)^2\simeq2\times 10^{-18}$ eV$^{-1}$ cm$^3$, 
and for this ratio of $\lambda/(mc^2)^2$, $\lambda\simeq2\times10^{-62}$ eV cm$^3$ 
when $m\simeq10^{-22}$ eV (see Refs. \cite{RS1}, and references therein). 
Therefore, SFDM provides $\lambda_{\rm{deB}}$ and $l_{\rm{SI}}$ as two mechanisms to 
suppress small-scale structures. When $l_{\rm{SI}}\gg\lambda_{\rm{deB}}$, 
only $l_{\rm{SI}}$ is responsible for affecting structure formation. 
This is the self-interaction-dominated limit, also known as the Thomas-Fermi regime; 
we called it TYPE II BEC-CDM in Ref. \cite{RS1}. We will also 
address the limit in which there is \textit{no} self-interaction (i.e. $\lambda\equiv 0$, 
also known as fuzzy dark matter (FDM) in Refs. \cite{HBG, WC2009}; we called it TYPE I BEC-CDM in Ref. \cite{RS1}).

This paper is organized as follows: in Section \ref{sec:EOM}, 
we present the fundamental equations underlying the description of SFDM with a quartic, 
positive self-interaction. In Section \ref{sec:background}, 
we solve for the homogenous background evolution of a universe 
with the same cosmic inventory as $\Lambda$CDM, 
but with CDM replaced by SFDM, over cosmic time. 
We identify three distinctive phases in the evolution of SFDM: 
non-relativistic, dust-like behavior at late times, which is indicative of the usefulness of SFDM 
as cold dark matter, a radiationlike phase at intermediate times, 
and an even earlier phase when SFDM behaves as a ``stiff'' relativistic fluid. 
We note that SFDM is relativistic in both the radiationlike phase and the stiff phase 
(in this work, the word ``relativistic'' does not only refer to radiation, but generally refers to 
any type of matter for which the ratio of pressure $p$ to energy density $\rho$ is 
in the physically allowed range $1/3\leq p/\rho \leq 1$). 
While the former two phases and the corresponding constraint 
from the time of matter-radiation equality at $z_{\rm{eq}}\sim3000$ have been identified and 
appreciated previously (e.g. in Refs. \cite{Goodman, Peebles2000}), 
the latter one has only been sporadically encountered, and often as 
a result of special assumptions; see, e.g. Refs. \cite{Joyce1997, ALS2002, Dutta2010}. 
However, we find that the stiff phase is generic for complex SFDM, 
no matter which values of SFDM parameter one adopts. We will comment more on this later. 
In Section \ref{sec:constraints}, we present the most important results of this work, 
namely the constraints on the SFDM model parameters, boson mass $m$, 
and \textit{positive} boson self-interaction coupling strength $\lambda$ 
(or equivalently $\lambda/(mc^2)^2$, in which the final results will actually be presented), 
which follow from the constraints on the homogeneous background evolution 
by current cosmological observations. These include the aforementioned redshift of 
matter-radiation equality $z_{\rm{eq}}$ and the effective number of neutrino species $N_{\rm{eff}}$ 
at the time of big bang nucleosynthesis (BBN). They constrain the timing and longevity of the stiff 
and radiationlike phases of SFDM, and thereby set severe restrictions on the allowed parameter space. 
Finally, Section \ref{sec:discussion} contains detailed discussions on the many implications of our results, 
while Section \ref{sec:conclusion} presents a brief summary. 
Appendices \ref{app:pfrw}-\ref{app:matching} contains some more technical aspects 
which have been deferred from the main text, but help to make the presentation more self-contained. 

In deriving those constraints on SFDM in concordance with current cosmological observations, 
we obtain three main results: 
First, we are able to restrict the allowed parameter space of SFDM \textit{severely}, 
despite the fact that we limit our consideration to the homogeneous background universe. 
Second, there, nevertheless, remains a semi-infinite stripe in parameter space 
which is in accordance with observations, including parameter sets 
which are able to resolve the small-scale problems of CDM. 
Third, the currently favored value of $N_{\rm{eff}}$ during BBN, 
which exceeds the standard value of 3.046 for a universe containing 
just three neutrino species and no extra relativistic species, 
excludes the possibility that the dark matter is SFDM 
with \textit{vanishing} self-interaction, i.e., fuzzy dark matter, at $>68\%$ confidence. 
On the contrary, SFDM with self-interaction provides a natural explanation 
of why $N_{\rm{eff}}$ during BBN \cite{Steigman2012} is higher than 
that inferred from the Cosmic Microwave Background \cite{Planck}.

\section{Basic equations}  \label{sec:EOM}

We will assume in this paper that dark matter is described by a
complex field. There are several motivations for considering a
complex, rather than a real field, namely the U(1) symmetry
corresponding to the dark matter particle number (charge) conservation 
(see Appendix \ref{app:charge} and Ref. \cite{BCK2002}), 
and the richer dynamics of halos, e.g. formation of vortices (see Refs. \cite{RS1, RS2}).

\subsection{Equation of motion for SFDM} \label{section:EOM1}

The ground state of a bosonic system can be described by a
classic scalar field theory. We choose the following generic
Lagrangian density of the complex scalar field
\begin{equation} \label{equation:lag}
    \mathscr{L}=\frac{\hbar^2}{2m}g^{\mu\nu}\partial_\mu\psi^*\partial_\nu\psi-V(\psi).
\end{equation}
The metric signature we adopt here is $(+, -, -, -)$. The potential
in the Lagrangian above contains a quadratic term accounting for the
rest-mass plus a quartic term accounting for the self-interaction
\begin{equation} \label{equation:potential}
    V(\psi)=\frac{1}{2}mc^2|\psi|^2+\frac{\lambda}{2}|\psi|^4.
\end{equation}
This model has been adopted in other works, as well; see, e.g., 
Refs. \cite{Goodman}, \cite{PV1999}, \cite{MM2012}. We choose physical units
throughout, in contrast to the convention usually used in
high-energy particle physics. The main reason is that this is the
first paper in a series of works on the cosmological behavior of
SFDM, which will include the linear and nonlinear growth of fluctuations.
There, we are concerned with non-relativistic ($c \to \infty$) and
classical limits ($\hbar \to 0$), where natural units become
disadvantageous. In order for $\mathscr{L}$ to have units of energy
density, the field has units of $[\psi] = \rm{cm}^{-3/2}$ and the
unit for the coupling constant is $[\lambda] =$ eV cm$^3$. A value of $\lambda = 2\times 10^{-62}$
eV cm$^3$ would correspond to $\hat \lambda = 2.6 \times 10^{-86}$. 
For the purpose of comparison, we take a look at the dimensionless self-interaction strength of 
QCD axions. According to equation (2) and (3) in Ref. \cite{SY2009}, 
$\hat \lambda_{\rm{axion}}\sim 10^{-53}$, also tiny, 
for the axion decay constant $f\simeq10^{12}$ GeV.

The quartic term in the above potential models the two-particle
self-interaction. It is a good approximation to ignore higher order
interactions when the bosonic gas is dilute, i.e., when the particle
self-interaction range is much smaller than the mean interparticle
distance. Moreover, since particles in non-zero-momentum states can be
neglected, it is sufficient to consider only two-body s-wave
scatterings. This means the coupling coefficient $\lambda$ is a
constant and related to the s-wave scattering length $a_s$ as
$\lambda=4\pi\hbar^2a_s/m$, which is effectively the first Born
approximation.

The equation of motion for the scalar field is the relativistic Klein-Gordon equation,
\begin{equation} \label{equation:relkg1}
    \frac{1}{\sqrt{-g}}\partial_\mu\left(g^{\mu\nu}\sqrt{-g}\partial_\nu\psi\right)+\frac{m^2c^2}{\hbar^2}\psi+\frac{2\lambda m}{\hbar^2}|\psi|^2\psi=0,
\end{equation}
or 
\begin{equation} \label{equation:relkg}
    g^{\mu\nu}\partial_\mu\partial_\nu\psi-g^{\mu\nu}\Gamma^\sigma_{~\mu\nu}\partial_\sigma\psi+\frac{m^2c^2}{\hbar^2}\psi+\frac{2\lambda m}{\hbar^2}|\psi|^2\psi=0,
\end{equation}
where $g_{\mu\nu}$ is the metric tensor and
$\Gamma^\sigma_{~\mu\nu}=\frac{1}{2}g^{\sigma\rho}(\partial_\mu
g_{\rho\nu}+\partial_\nu g_{\rho\mu}-\partial_\rho g_{\mu\nu})$ is
the Christoffel symbol, calculated in Appendix \ref{app:newtoniangauge} for
the perturbed Friedmann-Robertson-Walker (FRW) metric. Combining
such a metric in the conformal Newtonian gauge with the Klein-Gordon
equation (\ref{equation:relkg}) yields
\begin{displaymath}
    \left(1-2\frac{\Psi}{c^2}\right)\frac{\partial^2_t\psi}{c^2}-\left(1+2\frac{\Phi}{c^2}\right)\frac{\nabla^2\psi}{a^2}+
    \frac{3\ud a/\ud t}{c^2a}\partial_t\psi-
\end{displaymath}
\begin{displaymath}
    - \left(\partial_t\Psi+3\partial_t\Phi+6\frac{\ud a/\ud
    t}{a}\Psi\right)\frac{\partial_t\psi}{c^4}-
\end{displaymath}
\begin{equation} \label{equation:relkgfull}
    -\nabla(\Psi-\Phi)\cdot\frac{\nabla\psi}{c^2a^2}+\frac{m^2c^2}{\hbar^2}\psi+\frac{2\lambda m}{\hbar^2}|\psi|^2\psi=0.
\end{equation}
Here, $a$ denotes the scale factor of the expanding FRW universe,
and $\Psi$ and $\Phi$ are the perturbations to the otherwise
homogeneous metric (see Appendix \ref{app:pfrw}, where we
summarize some of the more technical, but otherwise known
 derivations).

\subsection{Einstein field equations} \label{section:efg}

The perturbed metric given by equation (\ref{equation:metric}) is
related to the total mass-energy density of the universe through the Einstein field equations. 
With the Ricci tensor calculated in Appendix \ref{app:efe}, let us
consider the contribution from the time-time component,
\begin{equation} \label{equation:tt}
    R^0_{~0}-\frac{1}{2}R=\frac{8\pi G}{c^4}T^0_{~0}.
\end{equation}
In fact, the left-hand side is
\begin{IEEEeqnarray}{rCl} \label{equation:g00}
    R^0_{~0}-\frac{1}{2}R & = & (1-2\Psi/c^2)R_{00}-R/2\nonumber\\
    & = & \frac{3(\ud a/\ud t)^2}{c^2a^2}+\frac{2\nabla^2\Phi}{c^2a^2}-\nonumber\\
   & & -\frac{6\ud a/\ud t}{c^4a}\left(\partial_t\Phi+\frac{\ud a/\ud t}{a}\Psi\right).
\end{IEEEeqnarray}
Thus, the time-time component (\ref{equation:tt}) becomes 
\begin{equation}\label{equation:pfriedmann}
	3\frac{(\ud a/\ud t)^2}{a^2}+2\frac{\nabla^2\Phi}{a^2}-6\frac{\ud a/\ud t}{c^2a}\left(\partial_t\Phi+\frac{\ud a/\ud t}{a}\Psi\right)=  \frac{8\pi G}{c^2}T^0_{~0}.
\end{equation}

We can evaluate the contribution of the scalar field to the energy-momentum
tensor, using the Lagrangian density in equation (\ref{equation:lag}) and
equation (\ref{stresstensor}), which yields
\begin{IEEEeqnarray}{rCl}\label{equation:sfdmtmn}
    T_{\mu\nu,~\rm{SFDM}} & = &\frac{\hbar^2}{2m}(\partial_\mu\psi^*\partial_\nu\psi+\partial_\nu\psi^*\partial_\mu\psi)-\nonumber\\
    & & -g_{\mu\nu}\left(\frac{\hbar^2}{2m}g^{\rho\sigma}\partial_\rho\psi^*\partial_\sigma\psi- \right.\nonumber\\
    & & \left. -\frac{1}{2}mc^2|\psi|^2-\frac{\lambda}{2}|\psi|^4\right).
\end{IEEEeqnarray}
Its time-time component is recognized as
\begin{IEEEeqnarray}{rCl} \label{equation:t00}
	T^0_{~0,~\rm{SFDM}} & = & \mathscr{H}=\frac{\hbar^2}{2mc^2}\left(1-2\frac{\Psi}{c^2}\right)|\partial_t\psi|^2+\nonumber\\
    	& & + \frac{\hbar^2}{2ma^2}\left(1+2\frac{\Phi}{c^2}\right)|\nabla\psi|^2+\nonumber\\
	& & + \frac{1}{2}mc^2|\psi|^2+\frac{1}{2}\lambda|\psi|^4.
\end{IEEEeqnarray}
where $\mathscr{H}$ is the Hamiltonian density of SFDM. Note
that $\mathscr{H}$ is not invariant under coordinate
transformations, because matter is coupled to the gravitational
field, hence the energy of the bosons is not conserved.

\section{Homogenous background universe}  \label{sec:background}

\subsection{Mass-energy content of the FRW
universe and the Friedmann equation}\label{sec:FRWbackground1}

In this paper, we will consider a universe with the same
cosmic inventory as the basic $\Lambda$CDM model except that 
CDM is replaced by SFDM (we will call it $\Lambda$SFDM model from now on). 
We will use the set of cosmological parameters from the recent Planck data release
\cite{Planck} (listed as basic in Table \ref{table:tab1}) to solve for the evolution of the homogeneous
background universe below. From those we derive some other cosmological 
parameters needed for the calculation. Note again that here $\Omega_{\rm{dm}}h^2$ 
refers to the present-day SFDM energy density instead of CDM. 
We will see later that SFDM indeed behaves as CDM at present. 
$\Omega_rh^2$ accounts for the ordinary radiation component, 
i.e., photons and the Standard Model neutrinos. For simplicity, the neutrinos are 
considered as \textit{massless} so that the total matter density fraction today is 
$\Omega_m=\Omega_b+\Omega_{\rm{dm}}$, where 
$\Omega_b$ stands for the baryon density fraction at present. The density fraction of 
the cosmological constant is $\Omega_\Lambda=1-\Omega_m-\Omega_r$. 

\begin{table}[h]
\begin{center}
{\renewcommand{\arraystretch}{1.5}
\renewcommand{\tabcolsep}{.3cm}
\begin{tabular}[t]{cc|cc}
Basic & ~ & Derived & ~\\
\hline
$h$ & 0.673 & $\Omega_mh^2$ & 0.14187\\
\hline
$\Omega_bh^2$ & 0.02207 & $\Omega_rh^2$ & $4.184\times 10^{-5}$\\
\hline
$\Omega_{\rm{dm}}h^2$ &  0.1198 & $z_{\rm{eq}}$ & 3390\\
\hline
$T_{\rm{CMB}}$/K & 2.7255 & $\Omega_\Lambda$ & 0.687\\
\end{tabular}
}\caption{Cosmological parameters. The values in the left column 
(`Basic') are quoted from the Planck collaboration: 
central values of the 68\% confidence intervals for the base $\Lambda\text{CDM}$
model with Planck+WP+highL data, see Table 5 in Ref. \cite{Planck}. We calculate
those in the right column (`Derived').}
\label{table:tab1}
\end{center}
\end{table}

The expansion of the homogeneous FRW universe is governed by 
the Friedmann equation, which is a special case of equation (\ref{equation:pfriedmann}),  
\begin{IEEEeqnarray}{rCl}\label{equation:friedmannbu}
    H^2(t) & \equiv & \left(\frac{\ud a/\ud t}{a}\right)^2\nonumber\\
    & = & \frac{8\pi G}{3c^2}\left[\bar{\rho}_r(t)+\bar{\rho}_b(t)+\bar{\rho}_\Lambda(t)+\bar{\rho}_{\rm{SFDM}}(t)\right],\IEEEeqnarraynumspace
\end{IEEEeqnarray}
where we have $\bar{\rho}_r(t) = \Omega_r\rho_{\rm{0,crit}}/a^4$ for radiation, $\bar{\rho}_b(t) =
 \Omega_b\rho_{\rm{0,crit}}/a^3$ for baryons, $\bar{\rho}_\Lambda(t) =
 \Omega_\Lambda\rho_{\rm{0,crit}}$ for the cosmological constant and 
 the SFDM energy density $\bar \rho_{\rm{SFDM}}(t)$ defined in the next section. 
 The critical energy density at the present epoch is 
 \begin{equation}
 	\rho_{\rm{0,crit}}=\frac{3H_0^2c^2}{8\pi G}.
 \end{equation}

Here is a technical detail: during the electron-positron annihilation 
that occurs around $0.5$ MeV, $\bar{\rho}_r$ does not simply evolve 
as $a^{-4}$ since photons get heated. Hence, we need to calculate 
the cosmic thermal history exactly, i.e., the photon temperature $T$ as a function of $a$ during that period, 
to acquire the evolution of $\bar{\rho}_r$. This effect will be reflected 
on the solutions in Section \ref{sec:FRWbackground4} 
(see Chapter 3 in Ref. \cite{Weinberg} for a standard treatment).
 
As for the SFDM, we will see in the next section that $\bar\rho_{\rm{SFDM}}$ 
evolves through three phases which can be characterized by different equations of state.

\subsection{Evolution of scalar field dark matter}
\label{sec:SFDMbackground}

In the case of the unperturbed homogeneous universe where
$\Psi=\Phi=0$, the scalar field is only a function of time, i.e., its
energy-momentum tensor is diagonal. Hence, SFDM can be treated as a
perfect fluid characterized by energy density $\bar \rho$, pressure
$\bar p$ and 4-velocity $u^\mu$ (for brevity, we omit the subscript SFDM in this section). The corresponding energy-momentum
tensor is
\begin{equation} \label{equation:fluidt}
    T_{\mu\nu}=(\bar \rho+\bar p)u_\mu u_\nu/c^2-g_{\mu\nu}\bar p,
\end{equation}
where $u^0=c$ and $u^i=0$ for the homogeneous background universe.
In fact, the energy density and pressure can be derived from
equations (\ref{equation:sfdmtmn}) and (\ref{equation:fluidt}),
\begin{IEEEeqnarray}{rCl}
    \bar{\rho} & = &T^0_{~0}=\frac{\hbar^2}{2mc^2}|\partial_t\psi|^2+\frac{1}{2}mc^2|\psi|^2+\frac{1}{2}\lambda|\psi|^4, \label{equation:densitybu}\\
    \bar{p} & = & -T^i_{~i}=\frac{\hbar^2}{2mc^2}|\partial_t\psi|^2-\frac{1}{2}mc^2|\psi|^2-\frac{1}{2}\lambda|\psi|^4 \label{equation:pressurebu}.
\end{IEEEeqnarray}

Without perturbation terms in equation (\ref{equation:relkgfull}), the equation of motion for homogeneous SFDM is then
\begin{equation} \label{equation:febu}
    \frac{\hbar^2}{2mc^2}\partial^2_t\psi+\frac{\hbar^2}{2mc^2}\frac{3\ud a/\ud t}{a}\partial_t\psi+\frac{1}{2}mc^2\psi+\lambda|\psi|^2\psi=0,
\end{equation}
It can be transformed into an equivalent form, namely the energy conservation
equation, given the expressions for $\bar{\rho}$ and $\bar{p}$ above,
\begin{equation} \label{equation:ec}
    \frac{\partial\bar{\rho}}{\partial t}+\frac{3\ud a/\ud t}{a}(\bar{\rho}+\bar{p})=0.
\end{equation}
Note that this is also one of the conservation laws of the energy-momentum tensor $T^{0\nu}_{~~;\nu}=0$, 
which is not surprising since the energy-momentum tensor is the Noether current 
of the spacetime translational symmetry and its conservation laws hold when the field follows 
the equation of motion (\ref{equation:febu}).

If there were an \textit{explicit} equation of state (EOS), relating $\bar p$
to $\bar \rho$, we could solve for the evolution of the entire background universe directly by
combining it with equation (\ref{equation:ec}) and the Friedmann equation (\ref{equation:friedmannbu}). 
As we show below, this is only possible in certain limits of $\bar w\equiv\bar p/\bar\rho$, 
but the SFDM will pass through these limits as it evolves. 
Hence, it will be instructive to identify these phases of its evolution first, 
before we solve the general evolution equation in detail.

One of the basic behaviors of a scalar field is oscillation over time \cite{Turner}, 
characterized by its changes in phase $\theta$. The oscillation angular frequency 
is defined as $\omega=\partial_t\theta$, the same as in Appendix \ref{app:charge}. 
We will see that the scalar field behaves differently whether $\omega$ 
predominates over the expansion rate $H$ or the contrary (oscillation vs. roll).

\subsubsection{Scalar field oscillation faster than Hubble expansion
($\omega/H\gg1$)} \label{sec:SFDMbackground1}

In this regime, the oscillation angular frequency can be derived as (see Appendix \ref{app:charge}) 
\begin{equation} \label{equation:dispersion}
    \omega=\frac{mc^2}{\hbar}\sqrt{1+\frac{2\lambda}{mc^2}|\psi|^2}.
\end{equation}
If $\omega$ is much larger than the Hubble expansion rate $H$, 
the exact cosmological time evolution of the scalar field will be hard to solve numerically, 
given that the necessary time step is essentially too tiny ($\propto1/\omega$). 
Instead, we follow the evolution of the time-average values of $\bar{\rho}$ and
$\bar{p}$ over several oscillation cycles. Multiplying the field equation 
(\ref{equation:febu}) by $\psi^*$ and then averaging over
a time interval that is much longer than the field oscillation
period, but much shorter than the Hubble time, results in 
(see Refs. \cite{Turner, PV1999}, and Appendix \ref{app:charge} for detailed derivation) 
\begin{equation}\label{equation:kineticvsother}
    \frac{\hbar^2}{2mc^2}\langle|\partial_t\psi|^2\rangle=\frac{1}{2}mc^2\langle|\psi|^2\rangle+\lambda\langle|\psi|^4\rangle.
\end{equation}
Combining this relation with the expressions for energy density and
pressure yields,
\begin{IEEEeqnarray}{rCl}
    \langle\bar{\rho}\rangle & = & mc^2\langle|\psi|^2\rangle+\frac{3}{2}\lambda\langle|\psi|^4\rangle\nonumber\\
    & \approx & mc^2\langle|\psi|^2\rangle+\frac{3}{2}\lambda\langle|\psi|^2\rangle^2,\label{equation:averagerho}\\
    \langle\bar{p}\rangle & = & \frac{1}{2}\lambda\langle|\psi|^4\rangle\approx\frac{1}{2}\lambda\langle|\psi|^2\rangle^2.
\end{IEEEeqnarray}
The equation of state is then
\begin{equation} \label{equation:eos1}
    \langle\bar{p}\rangle=\frac{m^2c^4}{18\lambda}\left(\sqrt{1+\frac{6\lambda\langle\bar{\rho}\rangle}{m^2c^4}}-1\right)^2, 
\end{equation}
 or equivalently,
\begin{equation}
  \langle\bar w\rangle \equiv \frac{\langle \bar p \rangle}{\langle \bar \rho
  \rangle} = \frac{1}{3}\left[\frac{1}{1+\frac{2mc^2}{3\lambda
  \langle |\psi|^2\rangle}}\right], 
\end{equation}
as found also in Ref. \cite{MUL2001} for a real scalar field. This equation of state (\ref{equation:eos1}) 
was also derived in Ref. \cite{Colpi1986}, in the context of boson stars. 
This approach will be called the \textit{fast oscillation approximation} in this paper.

\begin{enumerate}

\item[(1)] CDM-like phase: non-relativistic ($\langle\bar w\rangle=0$)
\label{sec:newtonlimit}

As the universe expands, the dark matter energy density 
will continuously decrease to the point when the
rest-mass energy density dominates the total SFDM energy density, i.e.,
$\frac{3}{2}\lambda\langle|\psi|^2\rangle^2\ll
mc^2\langle|\psi|^2\rangle$. In this limit, equation
(\ref{equation:eos1}) reduces to
\begin{equation}\label{equation:polytropiceos}
    \langle\bar{p}\rangle\approx\frac{\lambda}{2m^2c^4}\langle\bar{\rho}\rangle^2\approx 0,
\end{equation}
thus SFDM behaves like non-relativistic dust. Its self-interaction
is weak, so that on large scales SFDM is virtually collisionless.
Therefore, it evolves like CDM, following the familiar relation,
\begin{equation}
    \langle\bar{\rho}\rangle\propto a^{-3}.
\end{equation}
Then, the field amplitude decays as $|\psi| \propto a^{-3/2}$ 
and the scale factor goes as $a \sim t^{2/3}$.

\item[(2)] Radiation-like phase: relativistic ($\langle\bar w\rangle=1/3$)

At some point early enough, the SFDM will be so dense 
that the quartic term in the energy density (\ref{equation:averagerho}), the
self-interaction energy, dominates, i.e., 
$\frac{3}{2}\lambda\langle|\psi|^2\rangle^2\gg
mc^2\langle|\psi|^2\rangle$. In this limit, equation
(\ref{equation:eos1}) reduces to
\begin{equation}
    \langle\bar{p}\rangle\approx \frac{1}{3}\langle\bar{\rho}\rangle\approx\frac{1}{2}\lambda\langle|\psi|^2\rangle^2,
\end{equation}
thus the SFDM behaves like radiation. The time evolution is
accordingly
\begin{equation}
    \langle\bar{\rho}\rangle\propto a^{-4},
\end{equation}\\
while the field amplitude decays as $|\psi| \propto a^{-2}$ with the
scale factor $a \sim t^{1/2}$. 

It is important to note that SFDM without self-interaction, i.e., 
when $\lambda=0$, does \textit{not} undergo this
radiationlike phase. This has severe implications for such models,
as will be discussed in Section \ref{sec:fuzzy}.

\end{enumerate}

\subsubsection{Scalar field oscillation slower than Hubble expansion
($\omega/H\ll1$)} \label{sec:SFDMbackground2}

The Hubble parameter increases as one goes back in time, 
eventually exceeding the oscillation frequency, 
and the fast oscillation approximation will break down.
There is no simple explicit equation of state then.  
In this case, one has to solve the coupled equations (\ref{equation:friedmannbu}), 
(\ref{equation:densitybu}), (\ref{equation:pressurebu}) and (\ref{equation:ec}) exactly, 
with which we will be concerned in the next section. 
Nonetheless, one can still find a heuristic qualitative description, as follows.

\begin{enumerate}

\item[(1)] Stiff phase: relativistic limit ($\bar w=1$) 

At sufficiently early times, the expansion rate is much greater than
the oscillation frequency, $\omega/H\ll1$. The energy density and
pressure are both dominated by the first, kinetic term of
(\ref{equation:densitybu}) and (\ref{equation:pressurebu}),
for $(|\partial_t\psi|/|\psi|)^2\propto H^2$. Therefore,
\begin{equation} \label{stiff}
    \bar{p}\approx\bar{\rho}\approx\frac{\hbar^2}{2mc^2}|\partial_t\psi|^2.
\end{equation}
This stiff EOS implies that the sound speed almost reaches the speed
of light,  the maximal value possible, which is an analogue to the
incompressible fluid in Newtonian gas dynamics, where the sound
speed is infinity. In this case,
 \begin{equation}
    \bar{\rho}\propto a^{-6}, 
\end{equation}
and it can be shown that $\partial_t \psi \propto a^{-3}$, and hence $\psi \propto \log a$, 
where $a \sim t^{1/3}$. The physical picture of the stiff phase is that, 
at such an early epoch, the Hubble time is much smaller than the oscillation period 
so that the complex scalar field cannot even complete one cycle of spin, 
instead, it rolls down the potential well. The field value now evolves as $|\log a|$, 
which increases moderately compared with power laws as $a\to0$, 
suggesting that no undesirable blow-up occurs in this very early universe.

\end{enumerate}

\subsection{Evolution of the FRW homogeneous background universe with
SFDM} \label{sec:FRWbackground}

Now we are ready to calculate the full evolution history of the
homogeneous background universe, in which SFDM follows different 
equations of state (either explicit or implicit) at different cosmic epochs, 
while the other components can be treated straightforwardly 
as explained in Section \ref{sec:FRWbackground1}. 

\subsubsection{Numerical Method}\label{sec:method}

We have seen in Section \ref{sec:SFDMbackground1} that SFDM oscillates rapidly 
in comparison with the Hubble expansion rate at later times in the cosmic history. 
When $\omega/H\gg1$, the fast oscillation approximation can be applied, 
and we are able to use the equation of state (\ref{equation:eos1}) 
for the time-average SFDM energy density and pressure. 
From the energy conservation equation (\ref{equation:ec}), we see that 
as long as the oscillation is much faster than the rate at which the scale factor changes, 
the time evolution of the SFDM energy density should be quite smooth, with minute oscillation amplitude, 
since the oscillations in $\bar\rho_{\rm{SFDM}}$ and $\bar{p}_{\rm{SFDM}}$ cancel out 
through integration. Therefore, $\bar\rho_{\rm{SFDM}}$ should almost equal its 
time-average value $\langle\bar{\rho}_{\rm{SFDM}}\rangle$, 
which is even true in the real scalar field case \cite{Magana2012}. 
Furthermore, we can convert the energy conservation equation (\ref{equation:ec}) as follows, 
\begin{equation}
	\frac{\ud}{\ud a}\langle\bar{\rho}_{\rm{SFDM}}\rangle+\frac{3(\langle\bar{\rho}_{\rm{SFDM}}\rangle
	+\langle\bar{p}_{\rm{SFDM}}\rangle)}{a}=0, 
\end{equation}
so that it can be coupled to the equation of state (\ref{equation:eos1}) to solve for 
the evolution of $\langle\bar{\rho}_{\rm{SFDM}}\rangle$ and
$\langle\bar{p}_{\rm{SFDM}}\rangle$ as a function of scale factor $a$, 
by integrating from the present-day backwards to the point where $\omega/H=200$ 
(still well into the fast oscillation regime). We then solve the Friedmann equation 
(\ref{equation:friedmannbu}) with $\bar\rho_{\rm{SFDM}}$ replaced by 
$\langle\bar{\rho}_{\rm{SFDM}}\rangle$. The resulting time-average Hubble expansion rate 
$\langle H^2\rangle$ should be almost the same as its exact value, 
since $\bar\rho_{\rm{SFDM}}\simeq\langle\bar{\rho}_{\rm{SFDM}}\rangle$. 
The present-day values are inferred from Table \ref{table:tab1}. 
We will refer to the solution obtained above as the `late-time solution', during the period in which 
time-averages are excellent approximations to the exact values. 

At earlier times up to the big bang, the system has to be solved exactly, 
since $\omega/H$ decreases and the fast oscillation approximation becomes invalid. 
Combining equations (\ref{equation:densitybu}) and (\ref{equation:pressurebu}), 
the equation of state is implicitly given by the following coupled ordinary differential equations,
\begin{IEEEeqnarray}{rl} 
     & \partial_t(\bar{\rho}_{\rm{SFDM}}-\bar{p}_{\rm{SFDM}})\nonumber\\
     =~&B\sqrt{1+\frac{4\lambda}{m^2c^4}(\bar{\rho}_{\rm{SFDM}}-\bar{p}_{\rm{SFDM}})},\label{equation:defB}\\
     & \nonumber\\
     & \frac{\hbar^2}{2m^2c^4}\left(\partial_tB+\frac{3\ud a/\ud t}{a}B\right)\nonumber\\
     =~& 2\bar{p}_{\rm{SFDM}}-\frac{m^2c^4}{4\lambda}\times\nonumber\\
     & \times\left(\sqrt{1+\frac{4\lambda}{m^2c^4}(\bar{\rho}_{\rm{SFDM}}-\bar{p}_{\rm{SFDM}})}-1\right)^2,\label{equation:eos2}
\end{IEEEeqnarray}
where the auxiliary variable $B$ is defined as
$B\equiv mc^2\partial_t|\psi|^2$. We will refer to it as the `early-time solution'. One can verify that, 
if the left-hand side of equation (\ref{equation:eos2}) is zero, i.e., Hubble expansion is negligible, 
the equation of state reduces to the one in (\ref{equation:eos1}) 
in the limit $\omega/H\gg1$. We solve for the time-dependence of 
$\bar{\rho}_{\rm{SFDM}}$, $\bar{p}_{\rm{SFDM}}$ and scale factor $a$ 
by solving the combination of the Friedmann equation (\ref{equation:friedmannbu}),
the energy conservation equation (\ref{equation:ec}) along with (\ref{equation:defB}) 
and (\ref{equation:eos2}), using a fourth-order Runge-Kutta solver. 
The integration starts from the point where we cease to apply 
the fast oscillation approximation at $\omega/H=200$, as mentioned above, 
back to the big bang, in a way that it matches to the late-time solution. 
The matching is not trivial, since there are 3 variables in the late-time solution ($\langle\bar\rho_{\rm{SFDM}}\rangle$, $\langle\bar p_{\rm{SFDM}}\rangle$ and $a$) but 4 variables in the early-time solution ($\bar\rho_{\rm{SFDM}}$, $\bar p_{\rm{SFDM}}$, $a$ and $B$).  For details on the matching condition, see Appendix \ref{app:matching}.

\subsubsection{Numerical solution: evolution of the fiducial model}\label{sec:solution}

\begin{figure*}
\begin{minipage}[b]{0.5\linewidth}
      \centering\includegraphics[angle=90,width=8.5cm]{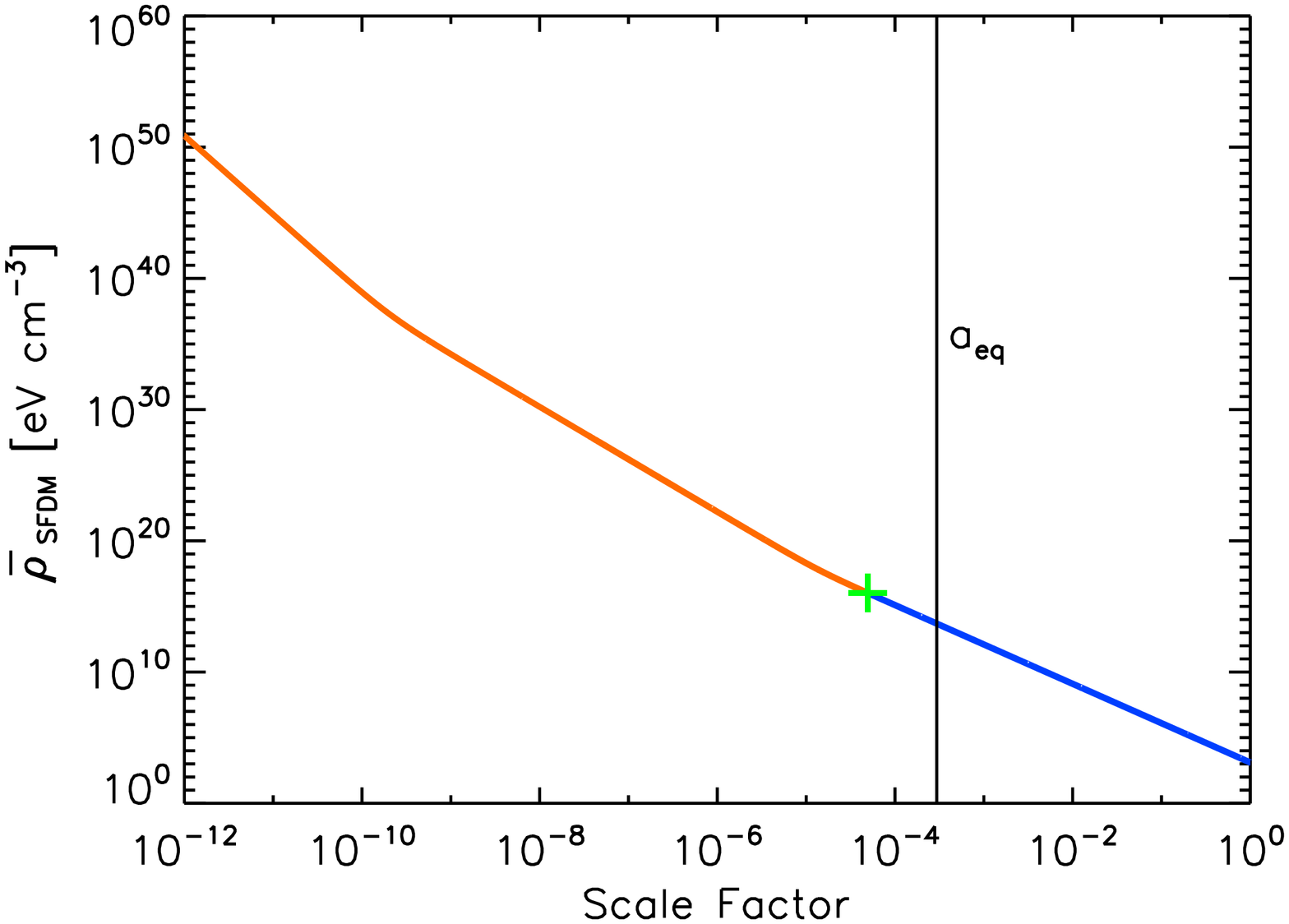}
     \hspace{0.1cm}
    \end{minipage}%
     \begin{minipage}[b]{0.5\linewidth}
      \centering\includegraphics[angle=90,width=8.5cm]{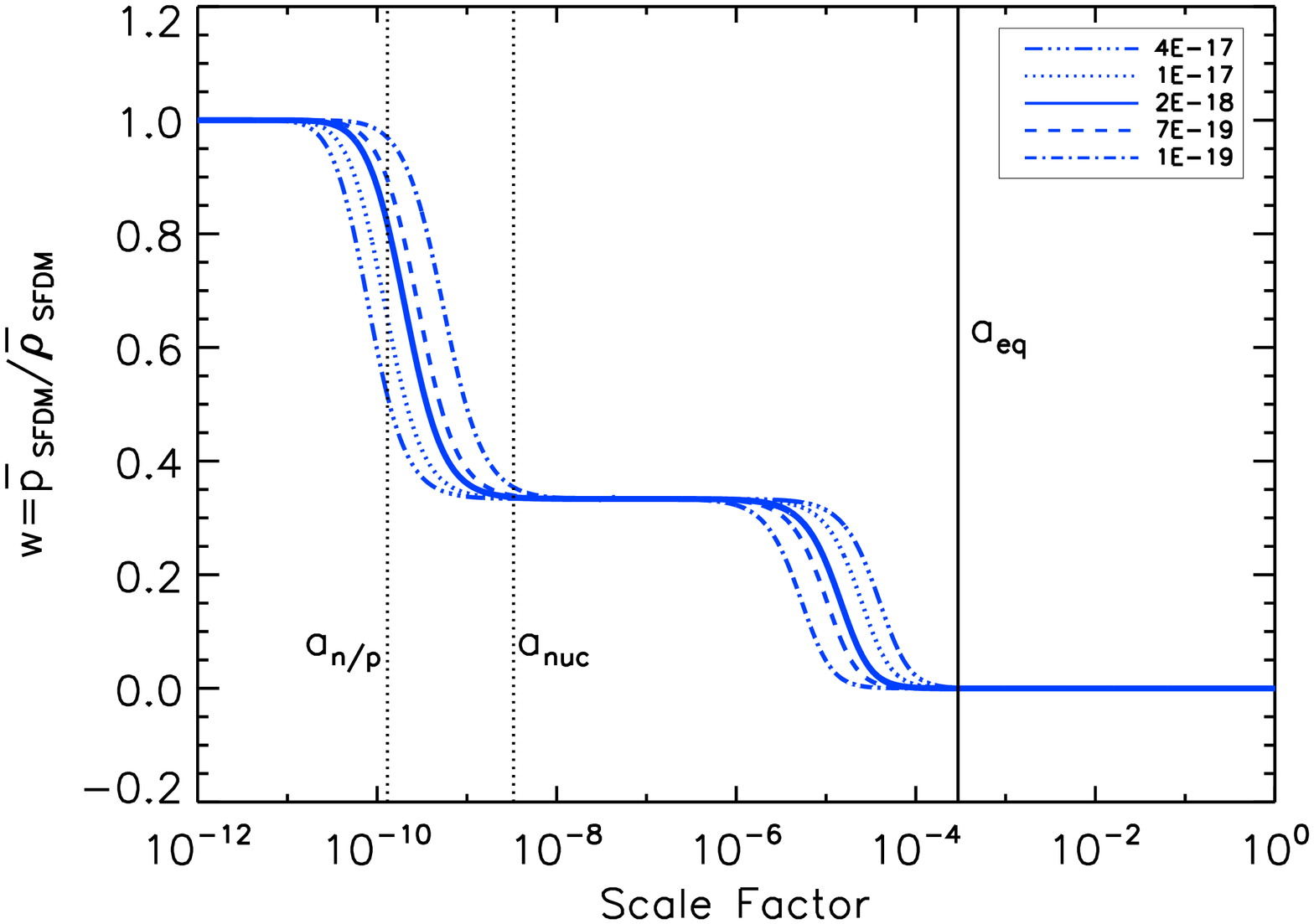}
     \hspace{0.1cm}
    \end{minipage}
 \caption{(Color online) \textit{Left-hand plot:} Evolution of the SFDM energy density $\bar\rho_{\rm{SFDM}}$ vs. scale factor $a$. The SFDM parameters are $m = 3\times10^{-21}$ eV$/c^2$ and $\lambda/(mc^2)^2 = 2\times10^{-18}~\rm{eV}^{-1}$ cm$^3$ (fiducial model).
 The vertical solid line depicts the epoch of matter-radiation equality $a_{\rm{eq}}$ from Table \ref{table:tab1}, while the cross indicates the point after
 which SFDM is well described as fully non-relativistic matter (CDM-like).
 \textit{Right-hand plot:} Evolution of the equation of state $\bar w = \bar{p}_{\rm{SFDM}}/\bar{\rho}_{\rm{SFDM}}$. The solid curve corresponds to the fiducial
 model plotted in the left panel. The other curves represent models with the same mass $m$, but different ratios of $\lambda/(mc^2)^2$ in unit of $\rm{eV}^{-1}$ cm$^3$, as seen in the
 legend.
 The vertical dotted lines depict the epoch of neutron-proton freeze-out $a_{\rm{n/p}}$ and the epoch of light-element production $a_{\rm{nuc}}$, respectively
 (see Section \ref{sec:FRWbackground4}). The larger the value of $\lambda/(mc^2)^2$, the longer lasts the radiationlike phase of SFDM:
 this provides constraints on this ratio from CMB observations of $a_{\rm{eq}}$ and $N_{\rm{eff}}$ during BBN, see Sections \ref{sec:FRWbackground3} and \ref{sec:FRWbackground4}. }
 \label{figure:EOSplots}
\end{figure*}

\begin{figure*}
\vspace{0.5cm}
    \begin{minipage}[b]{0.5\linewidth}
     \centering
     \includegraphics[angle=90,width=8.5cm]{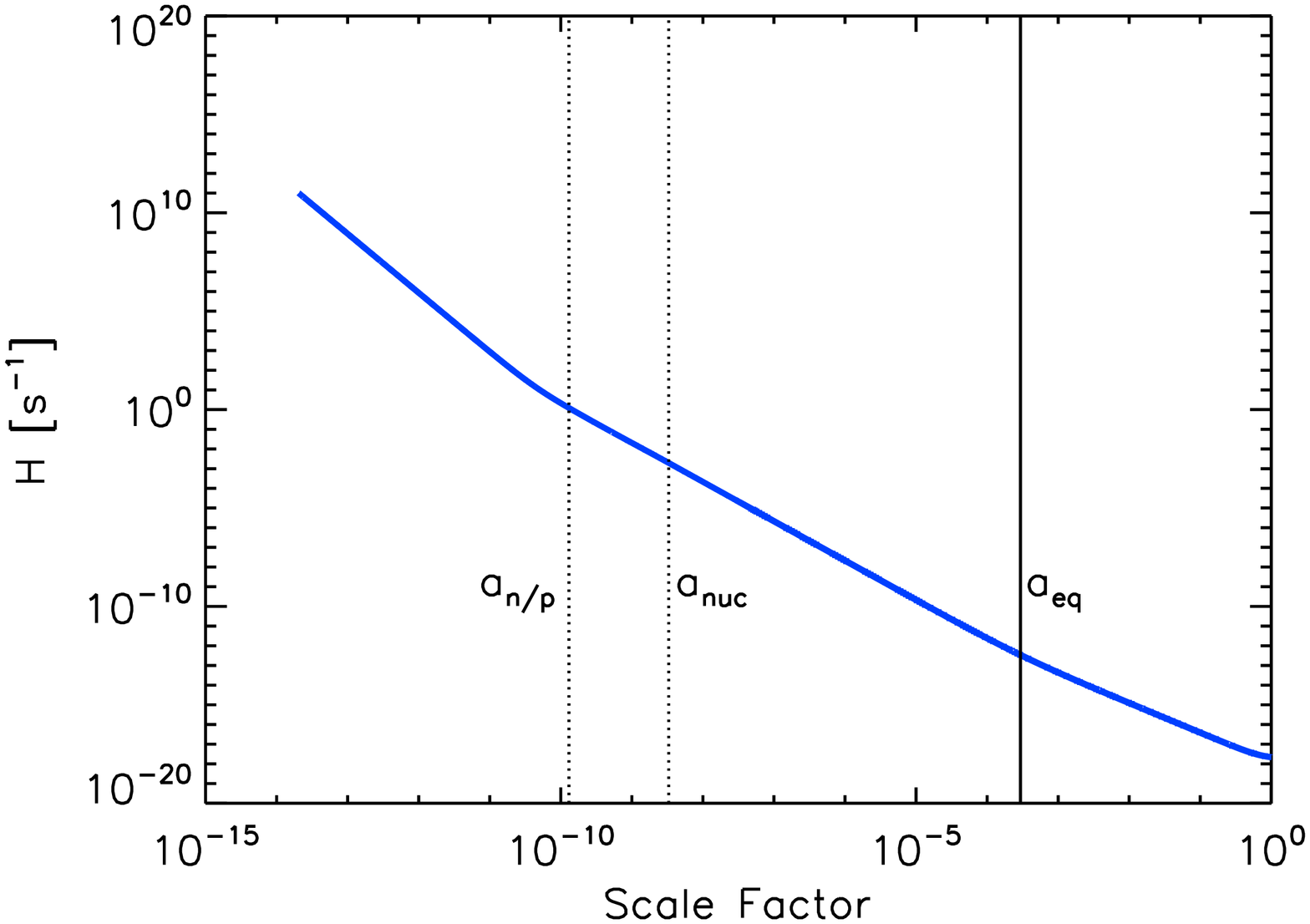}
     \hspace{0.1cm}
    \end{minipage}%
    \begin{minipage}[b]{0.5\linewidth}
      \centering\includegraphics[angle=90,width=8.5cm]{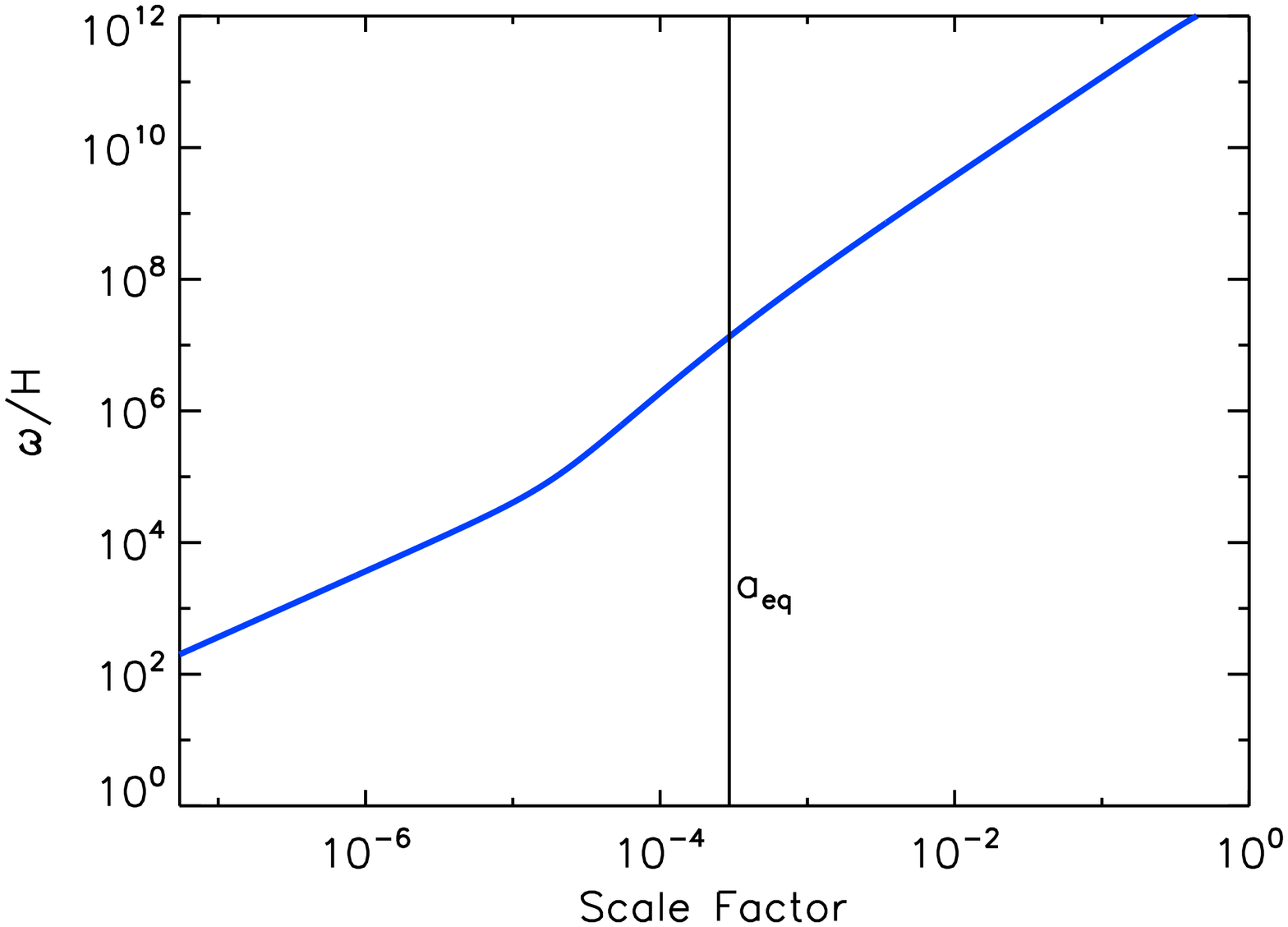}
     \hspace{0.1cm}
    \end{minipage}
 \caption{(Color online) \textit{Left-hand plot:} Hubble parameter $H(a)$ vs. scale factor $a$ for our fiducial SFDM model 
 with $m = 3\times10^{-21}$ eV$/c^2$ and $\lambda/(mc^2)^2 = 2\times10^{-18}~\rm{eV}^{-1}$ cm$^3$. 
 	\textit{Right-hand plot:} Evolution of the ratio of the oscillation angular frequency
 and Hubble parameter, $\omega/H$, for that same model. The vertical solid line depicts the epoch of matter-radiation equality $a_{\rm{eq}}$ from Table \ref{table:tab1}.
 The vertical dotted lines depict the beginning of the neutron-proton ratio freeze-out $a_{\rm{n/p}}$ and the epoch of light-element production $a_{\rm{nuc}}$, respectively
 (see Section \ref{sec:FRWbackground4}).}
 \label{figure:Hubble}
\end{figure*}

Anticipating our later results with regard to the cosmologically allowed
range of SFDM particle parameters, we will henceforth adopt the
following fiducial values for particle mass and self-interaction coupling strength: 
\begin{IEEEeqnarray}{l}
(m,\lambda)_{\rm{fiducial}} = (3 \times 10^{-21}~\rm{eV}/c^2, 1.8 \times 10^{-59}~\rm{eV~cm}^3),\nonumber\\
\lambda/(mc^2)^2=2\times 10^{-18}~\rm{eV^{-1}~cm^3}.
\end{IEEEeqnarray}
In this work, it is more convenient to work with the ratio $\lambda/(mc^2)^2$ rather than $\lambda$, 
as will be seen in the rest of the paper. The evolution for this fiducial SFDM model 
is shown in Figures \ref{figure:EOSplots} and \ref{figure:Hubble}. 
The smooth transition between the two parts of the solution (early-time and late-time) 
follows from the correctness of the matching conditions 
(see Appendix \ref{app:matching}). The evolution of the SFDM energy
density $\bar \rho_{\rm{SFDM}}$ in Figure \ref{figure:EOSplots}
(left-hand plot) shows not only the transition of SFDM from CDM-like
to radiationlike around $a\sim 10^{-4}$, but that at an even
earlier time $a \lesssim 10^{-10}$, SFDM follows, indeed, a stiff equation
of state. The evolution of the equation of state is plotted in
Figure \ref{figure:EOSplots} (right-hand plot), where we can also
clearly see the transition from the stiff phase, to the radiationlike 
phase, to the CDM-like phase.

\begin{figure*}
\begin{minipage}[b]{1\linewidth}
     \centering
     \includegraphics[angle=90,width=15cm]{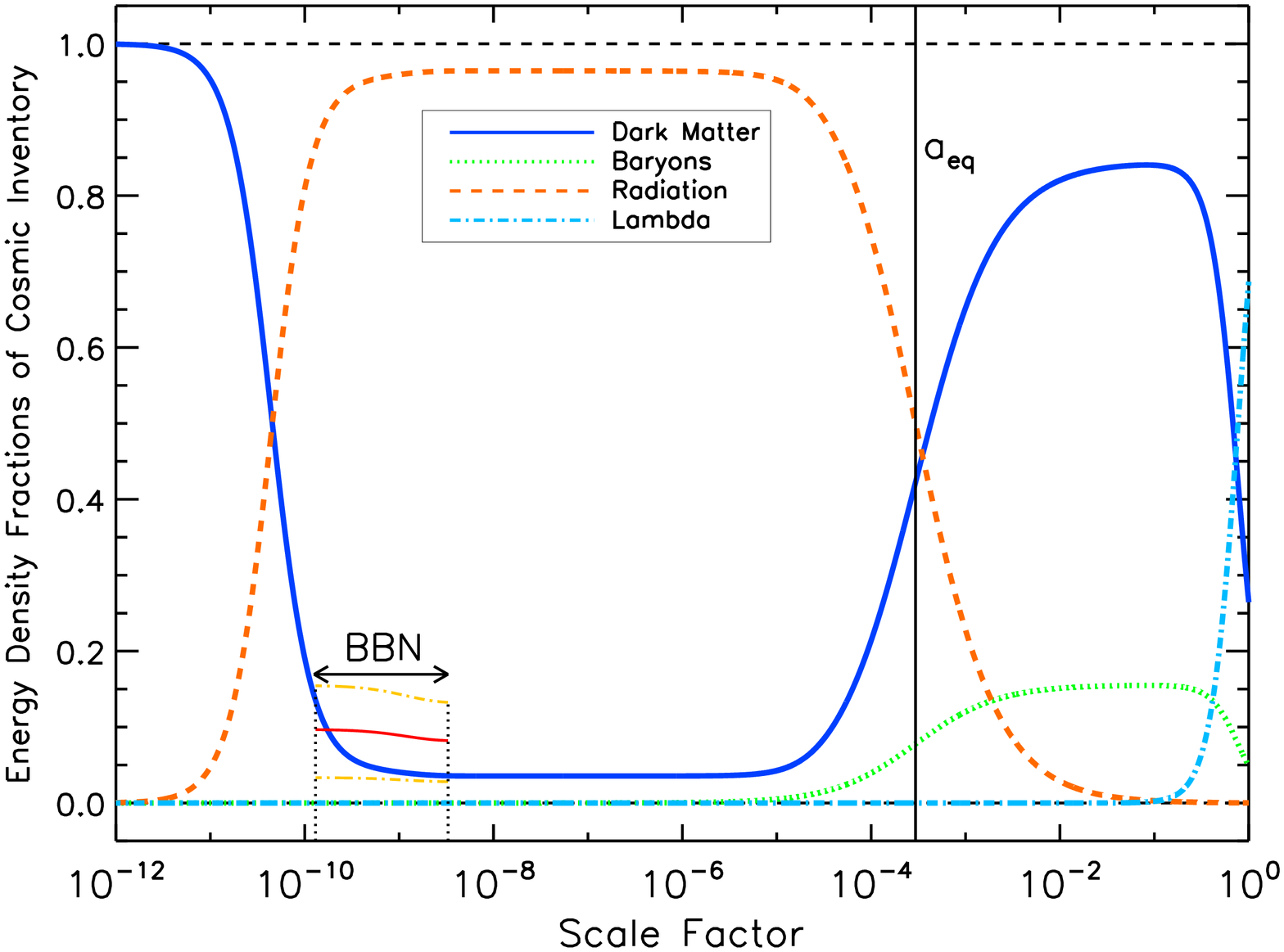}
     \vspace{0.5cm}
    \end{minipage}%
 \caption{(Color online) Evolution of the fractions $\Omega_i$ of the energy density of each cosmic component $i$ with
 SFDM of mass $m = 3\times10^{-21}$ eV$/c^2$ and self-interaction $\lambda/(mc^2)^2 = 2\times10^{-18}~\rm{eV}^{-1}$ cm$^3$ (fiducial model) represented by the thick curves. 
 Different components are depicted with different line styles, as labeled in the legend. 
 The solid vertical line corresponds to $a_{\rm{eq}}$. On the lower left part of the figure, 
 the thin curves represent the constraint from BBN. 
 The solid one refers to a universe with a constant $N_{\rm{eff}}$ of the central value in (\ref{equation:neff}) 
and the two dash-dotted ones refer to such universes with $N_{\rm{eff}}$ 
of the 1$\sigma$ limits there. The dotted vertical lines indicate 
the beginning of the neutron-proton ratio freeze-out $a_{\rm{n/p}}$ 
and the epoch of light-element production $a_{\rm{nuc}}$, respectively.
 }
 \label{figure:Omega}
\end{figure*}

The evolution of the energy content in our fiducial model can be found 
in Figure \ref{figure:Omega}. The energy density of SFDM
$\bar{\rho}_{\rm{SFDM}}\propto a^{-6}$ surpasses that of radiation
$\bar{\rho}_r\propto a^{-4}$ in the stiff phase of SFDM. Hence, the
expansion rate in the stiff phase is higher, $H\propto a^{-3}$, 
than that in the radiation-dominated era, $H\propto a^{-2}$. 
This is a ``scalar-field-dark-matter-dominated" era, before the radiation-dominated era. 
Here, the transition time from the stiff phase to the radiationlike phase 
depends on both $\lambda/(mc^2)^2$ \textit{and} $m$. 
This can be understood by realizing that, the transition happens
when the first term (kinetic term, which depends on $m$) and the third term
(self-interaction term, which depends on $\lambda$) 
on the rhs of (\ref{equation:densitybu}) and (\ref{equation:pressurebu}) become of equal order. 
Another way to see this is that, the equations which we solve 
when scalar field oscillation is slower than the Hubble expansion rate 
involve both these two parameters (see equations (\ref{equation:defB}) and  (\ref{equation:eos2})). 
After the stiff-to-radiation transition, the energy fraction of SFDM reaches a
``plateau'' as well as that of the regular radiation component,
since both components have radiationlike equations of state. 
This already implies that the kinetic term diminishes to the point 
where it is comparable to the self-interaction term (see equation(\ref{equation:kineticvsother})), 
as the scalar field oscillation becomes faster than the Hubble expansion rate, 
which is verified below. Therefore, the height of the plateau, 
i.e., the energy fraction of SFDM in the radiationlike phase,
is determined by $\lambda/(mc^2)^2$ alone, because the equations for
the fast oscillation approximation only concern this ratio (see equation
(\ref{equation:eos1})). It should be noted that the plateau height
would vanish if there is no self-interaction ($\lambda=0$), see also
Section \ref{sec:fuzzy}. 

The energy fraction of SFDM starts to rise from the plateau value
after a second transition from the radiationlike phase to the CDM-like phase. 
The energy density of SFDM evolves as $\bar{\rho}_{\rm{SFDM}}\propto a^{-3}$ 
like standard CDM, and the expansion rate as $H\propto a^{-3/2}$ when
SFDM dominates. The background evolution of the fiducial model is
then the same as the basic $\Lambda\rm{CDM}$ model.

It is interesting to note that, in the $\Lambda$SFDM model, dark matter dominates
over the other cosmological components \textit{twice} during the
cosmic history, first in the stiff-matter phase, where it is highly
relativistic, and later, when it behaves as pressureless dust, as in
the standard scenario of CDM. As we will see in the next section,
there are indeed constraints to be derived from both epochs. 
Also, the radiation-dominated era of the universe basically 
coincides with the radiationlike phase (plateau) of SFDM, 
since both of the SFDM transitions occur rapidly.

We would like to verify that the fast oscillation approximation 
discussed in Section \ref{sec:SFDMbackground1}
is indeed applicable for the fiducial model, for large enough $a$, 
where we solve for the evolution of the time-averages of $\bar{\rho}_{\rm{SFDM}}$ and $\bar{p}_{\rm{SFDM}}$, 
instead of solving for their exact values. In other words, 
we would like to see that its condition $\omega/H\gg1$ is fulfilled 
during that era, for our fiducial model. The plot of $\omega/H$ can be found in Figure \ref{figure:Hubble}
(right-hand plot). Apparently, $\omega/H>200$ for all $a$ therein, 
justifying the fast oscillation approximation at later times.

\section{Constraints on particle parameters from CMB and BBN measurements}\label{sec:constraints}

\subsection{Constraint from $z_{\rm{eq}}$} \label{sec:FRWbackground3}

As has been noted before (\cite{Goodman, Peebles2000, ALS2002}) the
transition of SFDM from the radiationlike phase to the CDM-like phase 
must happen early enough to be in agreement with the
redshift of matter-radiation equality $z_{\rm{eq}}$ determined by
the CMB temperature power spectrum, since its shape is subject to the
early integrated Sachs-Wolfe (ISW) effect, which depends upon
$z_{\rm{eq}}$ \cite{HuSugiyama1995}. In other words, in order to preserve $z_{\rm{eq}}$, 
SFDM should be well into the CDM-like phase at $z_{\rm{eq}}$. Before we proceed, 
it should be marked that the requirement above actually prohibits any freedom 
in choosing one of the initial conditions $\Omega_{\rm{dm}}h^2$, the present-day SFDM density parameter, 
which must be the same as that in the six-parameter 
base $\Lambda\rm{CDM}$ model (see Table \ref{table:tab1}). 
In fact, one can derive from the definition of $z_{\rm{eq}}$ that 
\begin{equation}
	1+z_{\rm{eq}}\equiv\frac{1}{a_{\rm{eq}}}=\frac{\Omega_bh^2+\Omega_{\rm{dm}}h^2}{\Omega_rh^2}, 
\end{equation}
where $a_{\rm{eq}}$ is the scale factor at matter-radiation equality. 
This justifies our choice of $\Omega_{\rm{dm}}h^2$.

The requirement that SFDM be fully non-relativistic at $z_{\rm{eq}}$ 
sets a constraint on the SFDM particle parameters, 
which is illustrated in Figure \ref{figure:EOSplots}. The redshift of 
matter-radiation equality $z_{\rm{eq}}$, according to Table \ref{table:tab1}, 
is marked as the vertical solid line in every plot. 
We define the cross in the left-hand plot 
to be the point at which $\langle\bar w\rangle\equiv\langle\bar p\rangle/\langle\bar\rho\rangle$ 
(neglecting the subscript SFDM here) is 0.001, 
a tiny deviation from zero, and 
consider SFDM after this point as fully non-relativistic. 
We can see that for the fiducial model, this point is 
indeed early enough compared with $z_{\rm{eq}}$. 
In fact, \emph{only} the ratio $\lambda/(mc^2)^2$ is constrained by this requirement, 
as it alone determines the radiation-to-matter
transition point of SFDM, resulting in 
\begin{equation} \label{equation:constraint1}
	\frac{\lambda}{(mc^2)^2} \leq 4 \times 10^{-17}~\rm{eV}^{-1}~\rm{cm}^3.
\end{equation}
This is the upper bound which would make the cross in the left-hand plot of 
Figure \ref{figure:EOSplots} lie on top of the vertical line indicating $z_{\rm{eq}}$, i.e., the
marginal case where SFDM has just fully morphed into CDM at matter-radiation equality 
(see also the right-hand plot for the evolution of $\langle\bar w\rangle$ in the marginal case). 
Equation (\ref{equation:constraint1}) implies that, 
even SFDM with large values of $\lambda$ \textit{and} $m$, as adopted in some literature, 
is able to fulfill this constraint (this is in the self-interaction-dominated limit, 
since large $m$ indicates small $\lambda_{\rm{deB}}$). 

The choice of the threshold 0.001 is artificial, though. If we relax it to 0.01, 
i.e., consider SFDM as fully non-relativistic when $\langle\bar w\rangle$ is less than 0.01, 
the corresponding constraint on $\lambda/(mc^2)^2$ would become 
$\lambda/(mc^2)^2\leq4.2 \times 10^{-16}~\rm{eV}^{-1}~\rm{cm}^3$, 
allowing a broader range of values. To determine this threshold, 
we need to calculate the CMB power spectrum for given SFDM 
particle parameters and see the range of them that preserves the early ISW effect. 
We plan this for future work.

\subsection{Constraint from $N_{\rm{eff}}$ during big bang nucleosynthesis}
\label{sec:FRWbackground4}

The abundances of the big bang nucleosynthesis (BBN) products set a
constraint on the Hubble expansion rate at that time, which depends
on the total energy density of the relativistic species,
parameterized by an effective number of relativistic degrees of
freedom, also known as an effective number of neutrino species,
$N_{\rm{eff}}$ (see Ref. \cite{Steigman2012} for a recent review). Thus,
measurements of the primordial abundance of helium and deuterium can
constrain the expansion rate or, equivalently, $N_{\rm{eff}}$, during
BBN. In the $\Lambda\rm{CDM}$ model, where there are only three SM 
neutrino species, $N_{\rm{eff, standard}}=3.046$ \cite{NEFFS}. 
In contrast, in $\Lambda$SFDM model, if SFDM is relativistic then, it will contribute to
$N_{\rm{eff}}$ as an extra relativistic component, and the
constraints on $N_{\rm{eff}}$ consequently put control on the
properties of SFDM, i.e., its particle parameters again.

The standard BBN scenario consists of two stages, the freeze-out of
the neutron fractional abundance and the production of light elements
combining free neutrons into nuclei, each affected by the expansion
rate at its own epoch. The attempts to determine $N_{\rm{eff}}$ from
BBN usually fit a cosmological model with constant
extra number of neutrino species $\Delta N_{\rm{eff}}\equiv N_{\rm{eff}}-N_{\rm{eff,standard}}$, 
e.g., with a constant portion of sterile neutrinos, to the primordial abundances of light elements 
extrapolated from observations. However, in $\Lambda$SFDM,  
the $\Delta N_{\rm{eff}}$ caused by SFDM
is changing over time as its equation of state varies during
different eras. Therefore, we must study the \emph{evolution} of
$N_{\rm{eff}}$ throughout BBN, which is an extended period from the
beginning of the neutron-proton ratio freeze-out around $T_{\rm{n/p}}=1.293$
MeV (the difference between the neutron and the proton mass) to the
epoch of nuclei production around $T_{\rm{nuc}}\approx0.07$ MeV.

In a $\Lambda$SFDM model, we infer the $N_{\rm{eff}}$ during BBN, namely from
$T_{\rm{n/p}}$ to $T_{\rm{nuc}}$, from the energy density of relativistic 
SFDM $\bar{\rho}_{\rm{SFDM}}$, which is determined by the particle parameters. 
In fact, SFDM is completely relativistic then and is the only source for $\Delta N_{\rm{eff}}$, 
\begin{equation}\label{equation:defneff}
	\frac{\Delta N_{\rm{eff}}}{N_{\rm{eff, standard}}}= \frac{\bar{\rho}_{\rm{SFDM}}}{\bar{\rho}_\nu},
\end{equation}
where $\bar{\rho}_\nu$ is the total energy density of the SM neutrinos. 
We compare the $N_{\rm{eff}}$ obtained this way to the 
measured value (constant over time) and impose a
conservative constraint that the $N_{\rm{eff}}$ during BBN be all
the time within 1$\sigma$ of the measured value,
\begin{equation} \label{equation:neff}
	N_{\rm{eff}}=3.71^{+0.47}_{-0.45},
\end{equation}
which we adopt from Ref. \cite{Steigman2012}. We shall adopt this 68\% confidence interval 
in constraining the parameters of SFDM in what follows. We note that while 
the standard $\Lambda$CDM model with $N_{\rm{eff, standard}}=3.046$ 
is inconsistent with the 1$\sigma$ constraint, it is, nevertheless, consistent within 95\% confidence. 
Ideally, we need to fit our model not to such a constant $N_{\rm{eff}}$ value, 
but to the data of primordial abundances directly by deriving those 
for $\Lambda$SFDM with a BBN code, which is intended as our future work. 

\begin{figure*}
\begin{minipage}[b]{0.5\linewidth}
      \centering\includegraphics[angle=0,width=8.5cm]{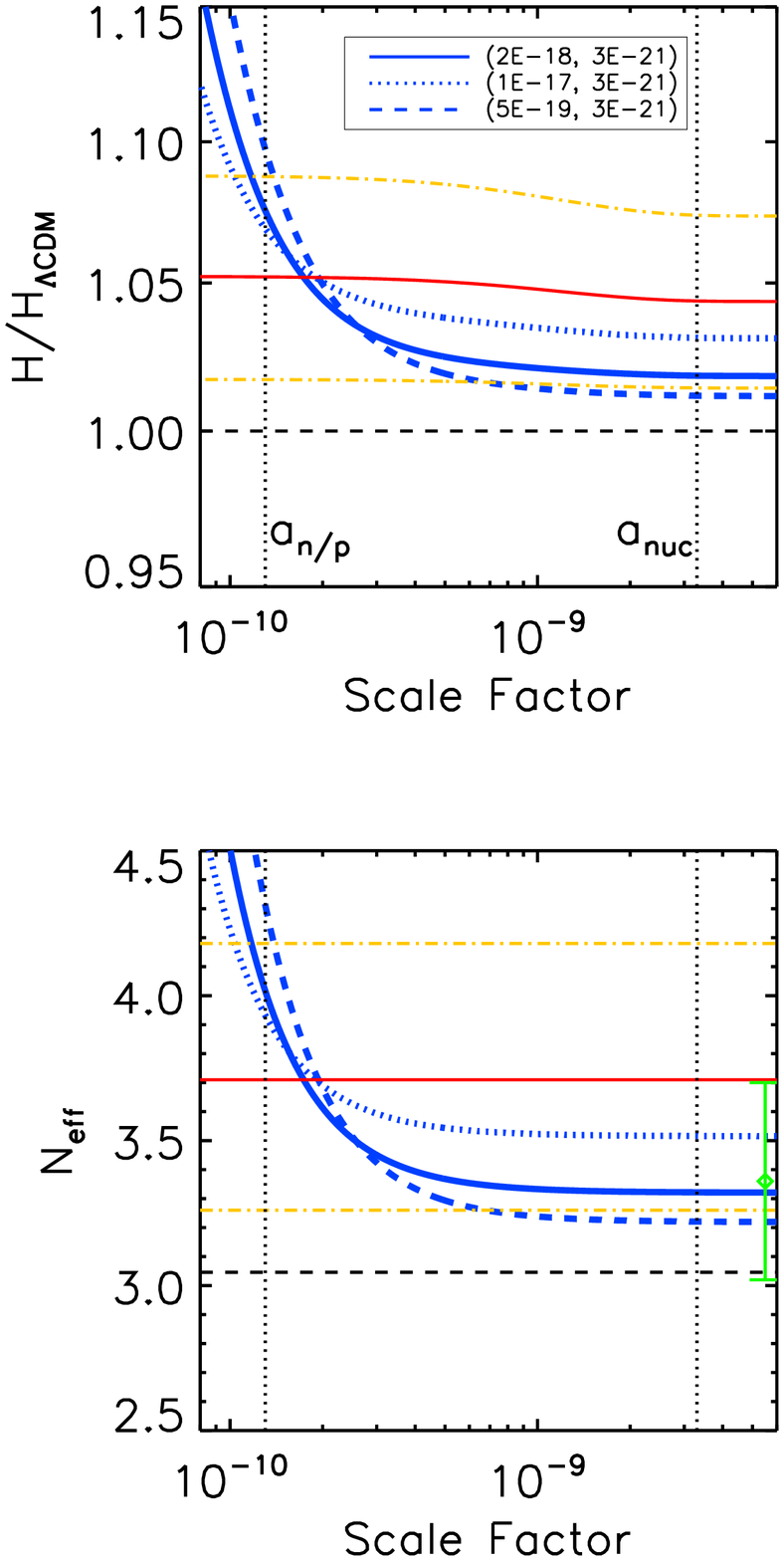}
     \hspace{0.1cm}
    \end{minipage}%
     \begin{minipage}[b]{0.5\linewidth}
      \centering\includegraphics[angle=0,width=8.5cm]{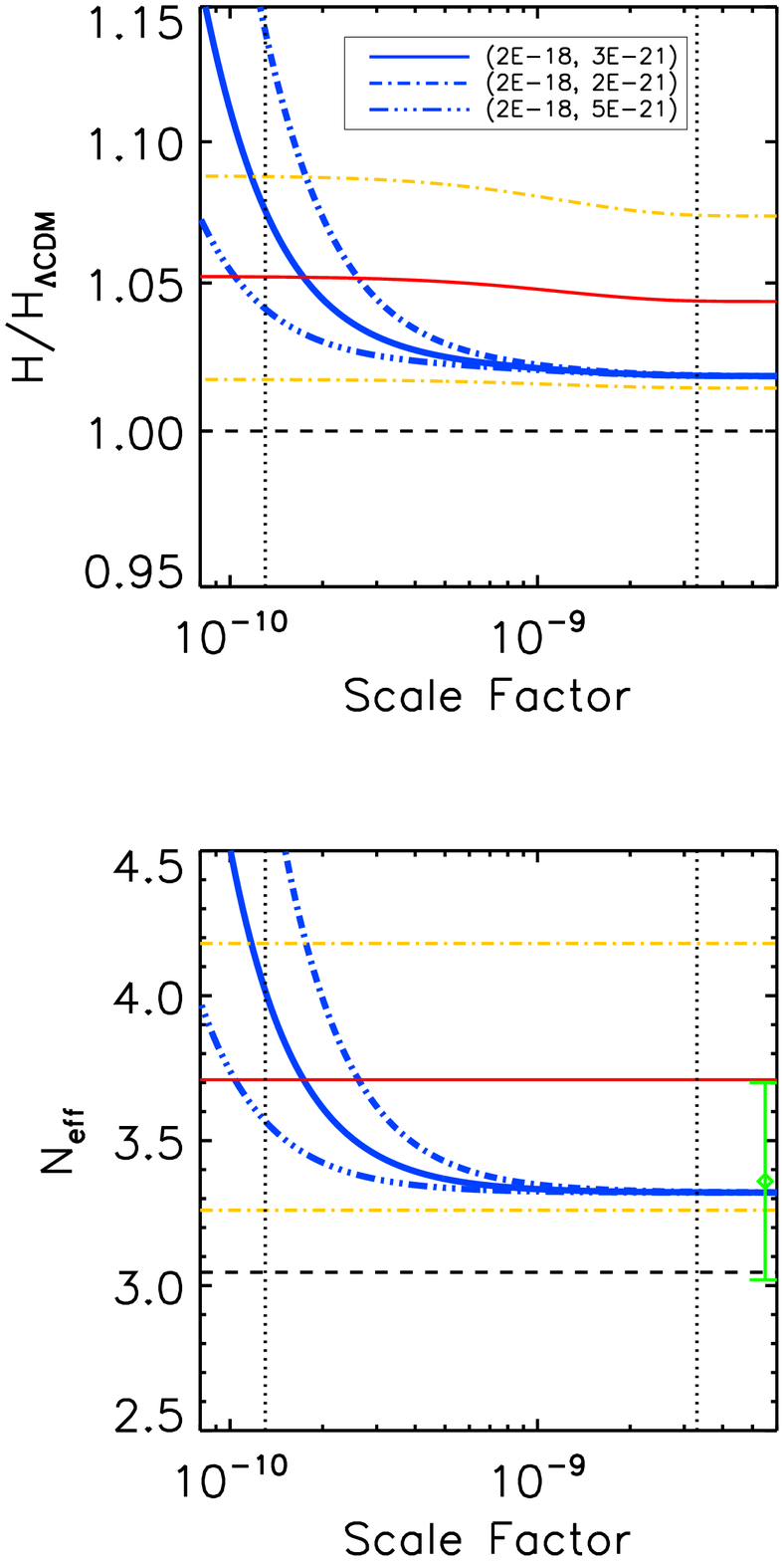}
     \hspace{0.1cm}
    \end{minipage}
 \caption{(Color online) 
 \textit{Upper plots:} Evolution of the normalized Hubble expansion rate $H(a)/H_{\Lambda\rm{CDM}}(a)$ vs. scale factor $a$.
 \textit{Lower plots:} Evolution of the effective number of neutrino species, $N_{\rm{eff}}$, vs. scale factor $a$. 
 The thick curves represent the evolution of $\Lambda$SFDM models with various particle parameters. 
 In the left-hand plots, $m$ is fixed. In the right-hand plots, $\lambda/(mc^2)^2$ is fixed. 
 The solid ones again correspond to our fiducial model with SFDM parameters $m = 3\times10^{-21}$ eV$/c^2$ and $\lambda/(mc^2)^2 = 2\times10^{-18}$ eV cm$^3$, see legends for the corresponding values of $(\lambda/(mc^2)^2, m)$ of each thick curve, in units of ($\rm{eV^{-1}~cm^3}$, eV/$c^2$). 
 Among the thin curves, the solid (dash-dotted) ones refer to universes with constant $N_{\rm{eff}}$ at the central value (68\% confidence limits) of the measured $N_{\rm{eff}}$ (\ref{equation:neff}). 
 The error bar in the lower plots is from the result of CMB measurements, $N_{\rm{eff}}=3.36\pm0.34$ \cite{Planck}.
 }
 \label{figure:Comparison}
\end{figure*}

The result is plotted in Figure \ref{figure:Comparison}. 
The upper plots show the Hubble expansion rate 
of $\Lambda$SFDM universes with different particle parameters
normalized to the expansion rate of the basic $\Lambda\rm{CDM}$
universe, which is an equivalent illustration of the evolution of
$N_{\rm{eff}}$, as in the lower plots. The thin curves are benchmarks. 
The solid ones refer to a universe with a constant $N_{\rm{eff}}$ of the central value 
in equation (\ref{equation:neff}) and the dash-dotted ones refer to such universes 
with $N_{\rm{eff}}$ of the 1$\sigma$ limits there, respectively. 
Note that in the upper plots for the normalized expansion rate, 
these thin curves are not straight lines due to the electron-positron annihilation. 
After this event, the neutrinos contribute less to the total energy density of the universe 
as their energy density fraction shrinks, because they are decoupled and do not get heated. 

In each plot, the thick curves denote different models of ($\lambda/(mc^2)^2$, $m$), 
according to the legend. The solid ones represent the fiducial model again: 
it complies with the constraint mentioned above (\ref{equation:neff}). 
It can be seen that these curves all reach the ``plateau'', 
i.e., the radiationlike phase, before the epoch of light-element production $a_{\rm{nuc}}$. 
The plateau height is purely 
determined by $\lambda/(mc^2)^2$, as explained in Section \ref{sec:solution}. 
In the left-hand plots, where we fix $m$, 
the higher the $\lambda/(mc^2)^2$, the higher the plateau. 
Meanwhile, earlier at $a_{\rm{n/p}}$ 
the transition from the stiff phase to the radiationlike phase 
may not have finished and the value of $N_{\rm{eff}}$ can be higher 
than its plateau, which is a function of both $\lambda/(mc^2)^2$ and $m$. 
In the right-hand plots, models with the same $\lambda/(mc^2)^2$, but different $m$, 
have the same plateau height, but diverge with a different rate as we go back in time: 
the lower the $m$, the later is the transition to the radiationlike phase.  
Therefore, the evolution of $N_{\rm{eff}}$ during BBN restricts both SFDM particle parameters, 
$\lambda/(mc^2)^2$ and $m$. This constraint is demonstrated in 
the next section and Figure \ref{figure:PPS}.

Note that this constraint is also illustrated in Figure \ref{figure:Omega}, where the definitions of the thin curves 
between $a_{\rm{n/p}}$ and $a_{\rm{nuc}}$, among which one is solid and two are dash-dotted, 
are the same as above, 
and the fraction of the SFDM energy density $\Omega_{\rm{SFDM}}$ is restricted by the two dash-dotted curves, 
which correspond to the 1$\sigma$ limits of $N_{\rm{eff}}$ in equation (\ref{equation:neff}). 
Again, these thin curves, which represent the energy fractions of extra radiation 
in models with constant $N_{\rm{eff}}$, slightly drop because of the electron-position annihilation. 
While $N_{\rm{eff}}$ characterizes the SFDM energy density (see equation (\ref{equation:defneff})), the relation between $\Omega_{\rm{SFDM}}$ and $N_{\rm{eff}}$ 
has a simple analytical form \textit{during the plateau}. The total energy density of a $\Lambda$SFDM universe during the radiation-dominated era is proportional to 
\begin{IEEEeqnarray}{rl}
     & 2+2N_{\rm{eff}}(\rm{plateau})\cdot\frac{7}{8}\left(\frac{4}{11}\right)^{4/3}\nonumber\\
    =~ & \left(2+2N_{\rm{eff,standard}}\cdot\frac{7}{8}\left(\frac{4}{11}\right)^{4/3}\right)\times\nonumber\\
    & \times\frac{1}{1-\Omega_{\rm{SFDM}}(\rm{plateau})}.
\end{IEEEeqnarray}
Thus, if SFDM reaches the plateau before $a_{\rm{nuc}}$, 
the 68\% confidence interval of $N_{\rm{eff}}$ (\ref{equation:neff}) can be converted to that of $\Omega_{\rm{SFDM}}$ during the plateau (its plateau height), using the equation above, 
\begin{equation}\label{equation:osfdm}
	0.028\le\Omega_{\rm{SFDM}}(\rm{plateau})\le 0.132.
\end{equation}
Consequently, we can use either (\ref{equation:neff}) or (\ref{equation:osfdm}) 
to constrain the SFDM parameter $\lambda/(mc^2)^2$, in terms of the plateau height,  
of those models in which SFDM has reached the radiationlike phase by the end of BBN. The result is
\begin{equation}\label{equation:lm2bound}
	9.5\times 10^{-19}~\rm{eV^{-1}~cm^3} \le\lambda/(mc^2)^2\le 1.5\times 10^{-16}~\rm{eV^{-1}~cm^3}, 
\end{equation}
as will be seen in Figure \ref{figure:PPS}. It should be also heeded that, in principle, SFDM 
does not have to reach the plateau by $a_{\rm{nuc}}$, and the result above (\ref{equation:lm2bound}) is not applicable for those models. 

\subsection{Result: allowed SFDM particle parameter space}
\label{sec:PPS}

\begin{figure*}
\begin{minipage}[b]{1\linewidth}
      \hspace{-1.5cm}
      \includegraphics[angle=90,width=16cm]{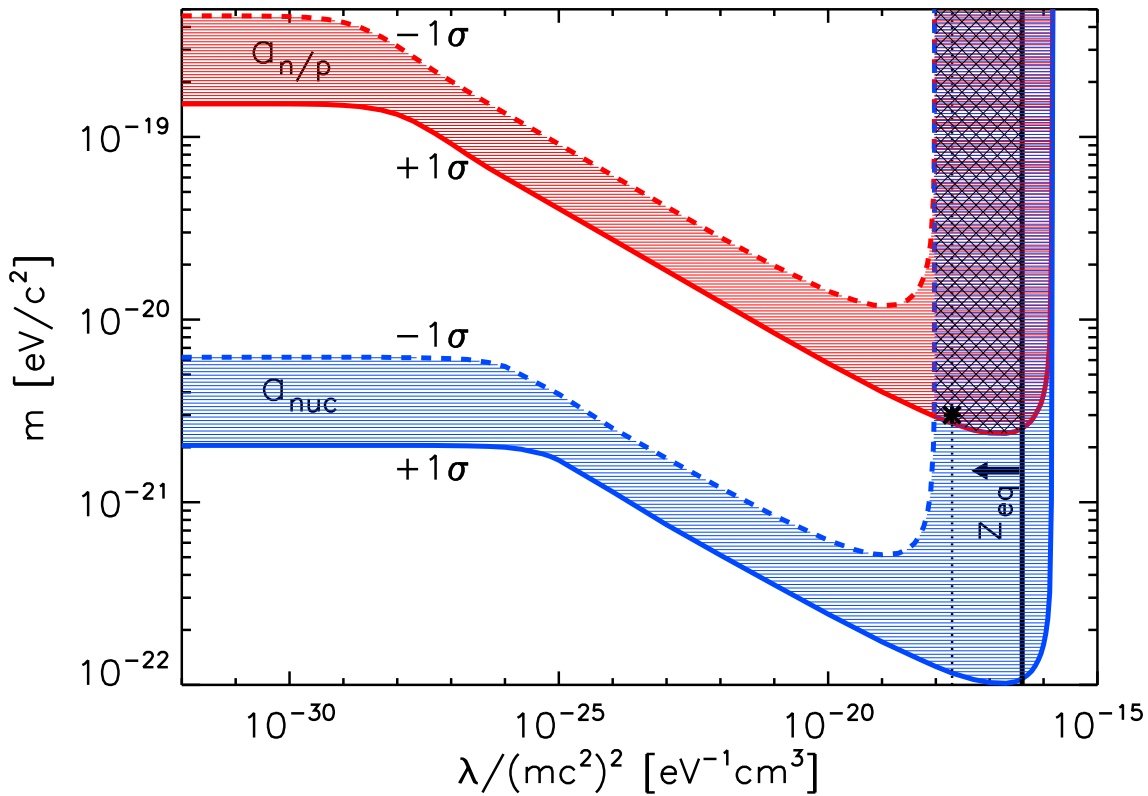}
     
    \end{minipage}%
 \caption{(Color online) 
 Parameter space of SFDM ($\lambda/(mc^2)^2$, $m$). The solid vertical line represent the upper bound 
 on $\lambda/(mc^2)^2$ which makes SFDM complete its transition from radiationlike to CDM-like 
 (i.e. $\langle\bar w\rangle=0.001$) just before the observed $z_{\rm{eq}}$. 
 The arrow indicates that the region on the left side of the solid vertical line 
 is allowed by this constraint from $z_{\rm{eq}}$. The two shaded bands are the allowed 
 regions derived from the constraints that $N_{\rm{eff}}$ be within the 1$\sigma$ interval of the value 
 measured by BBN, at $a_{\rm{n/p}}$ and $a_{\rm{nuc}}$, as labeled respectively. 
 For each band, the thick solid (dashed) boundary curve corresponds to the upper (lower)
1$\sigma$ limit of the measured value of $N_{\rm{eff}}$ in equation (\ref{equation:neff}).
 The final allowed region is crosshatched, after combining all constraints. 
 Our fiducial model, indicated by the star at $m=3\times10^{-21}\rm{eV}/c^2$, 
 lies on the dotted vertical line at $\lambda/(mc^2)^2=2\times10^{-18}~\rm{eV^{-1}~cm^3}$ 
 which corresponds to a radius of an equilibrium halo around 1 $\rm{kpc}$ .
 }
 \label{figure:PPS}
\end{figure*}

Combining the results from the above two sources of constraints, we
can confine the allowed region in the parameter space of SFDM, or
ultralight bosonic particle, see Figure \ref{figure:PPS} for the
parameter space plot. The constraint from $z_{\rm{eq}}$ is given by
the solid vertical line: the region on its left side is allowed, 
as shown by equation (\ref{equation:constraint1}). For the
constraint from $N_{\rm{eff}}$ during BBN, we sample the parameter
space to obtain the critical parameter values which marginally
fulfill the $1\sigma$ limits (\ref{equation:neff}). The two shaded bands correspond to the constraints that
$N_{\rm{eff}}$ be within 1$\sigma$ at $a_{\rm{n/p}}$ and $a_{\rm{nuc}}$, as labeled respectively. 
For each band, the thick solid (dashed) boundary curve refers to the upper (lower)
1$\sigma$ limit of $N_{\rm{eff}}$. The intersection of these two
bands represents the range of parameters that is consistent with the
$N_{\rm{eff}}$ constraint within $1\sigma$ throughout BBN. It is easily seen from the figure that 
all allowed choices of  ($\lambda/(mc^2)^2$, $m$) from the $N_{\rm{eff}}$ constraint 
indeed correspond to models in which SFDM has reached 
the radiationlike phase by the end of BBN, 
so that $\lambda/(mc^2)^2$ must be bounded within the asymptotic vertical lines 
(\ref{equation:lm2bound}) explained in the last section. 
This fact is completely due to the present-day measured $N_{\rm{eff}}$ value 
(\ref{equation:neff}). Should the 68\% confidence interval of $N_{\rm{eff}}$ be broaden, 
models in which SFDM had not reached the plateau by the end of BBN might also be allowed. 
Such models would not lie within the asymptotic vertical bounds of $\lambda/(mc^2)^2$ 
in the parameter space, as mentioned at the end of the last section. 

The final allowed region is given by combining all the constraints,
leaving the crosshatched area. The dotted vertical line, where
the fiducial model sits, has the value
$\lambda/(mc^2)^2=2\times10^{-18}~\rm{eV^{-1}~cm^3}$, which
corresponds to models with parameters for an equilibrium halo of
size about 1 $\rm{kpc}$, see equation (\ref{equation:polytroperadius}) in Section \ref{sec:coresize}. 
We can see that it lies within the allowed region, for high enough particle mass $m$. 
The significance of this result will be discussed in Section \ref{sec:coresize}.

\section{discussion}\label{sec:discussion}

\subsection{Relation between $N_{\rm{eff}}$ and smallest dark matter structure}
\label{sec:coresize}

We mentioned in the introduction that standard CDM meets 
challenges on small scales (mainly the cuspy core problem and the missing satellites problem), 
which could be possibly resolved if dark matter clustering 
is prohibited below certain scales. As a matter of fact, it has been pointed out in 
previous literature, e.g., Refs. \cite{Goodman, BoehmerHarko2007, RS1}, that 
self-interacting SFDM implies a minimum length scale $\sim l_{\rm{SI}}$ for a virialized object.
Though it is negligible compared with the energy density, as we pointed out in Section \ref{sec:newtonlimit} 
(i.e., SFDM behaves as collisionless dust on large scales), this self-interaction pressure 
affects the dynamics of small-scale nonlinear structures in the dark matter, 
just as thermal gas pressure does for the baryons.

In fact, equation (\ref{equation:polytropiceos}) is an $n=1$ 
polytropic equation of state $p\propto\rho^2$,
whose coefficient is proportional to $\lambda/(mc^2)^2$. This is true even 
for the inhomogeneous case, if we replace the background $\langle\bar p_{\rm{SFDM}}\rangle$ 
and $\langle\bar\rho_{\rm{SFDM}}\rangle$ by local values. Therefore, 
the minimum length scale in the self-interaction-dominated limit 
is then given by the radius of a virialized $n=1$ polytrope
\begin{equation}\label{equation:polytroperadius}
	 R=\pi\sqrt{\frac{\lambda}{4\pi Gm^2}}=\pi c^2\sqrt{\frac{\lambda}{4\pi G(mc^2)^2}},
\end{equation}
which is a function of $\lambda/(mc^2)^2$ only \cite{Goodman}. Note that $R\propto l_{\rm{SI}}$ 
up to a factor of order unity, and it is more precise to use $R$ for purposes 
with regard to a virialized dark matter halo.

\begin{figure*}
\begin{minipage}[b]{1\linewidth}
      \hspace{-1.5cm}
      \includegraphics[angle=90,width=16cm]{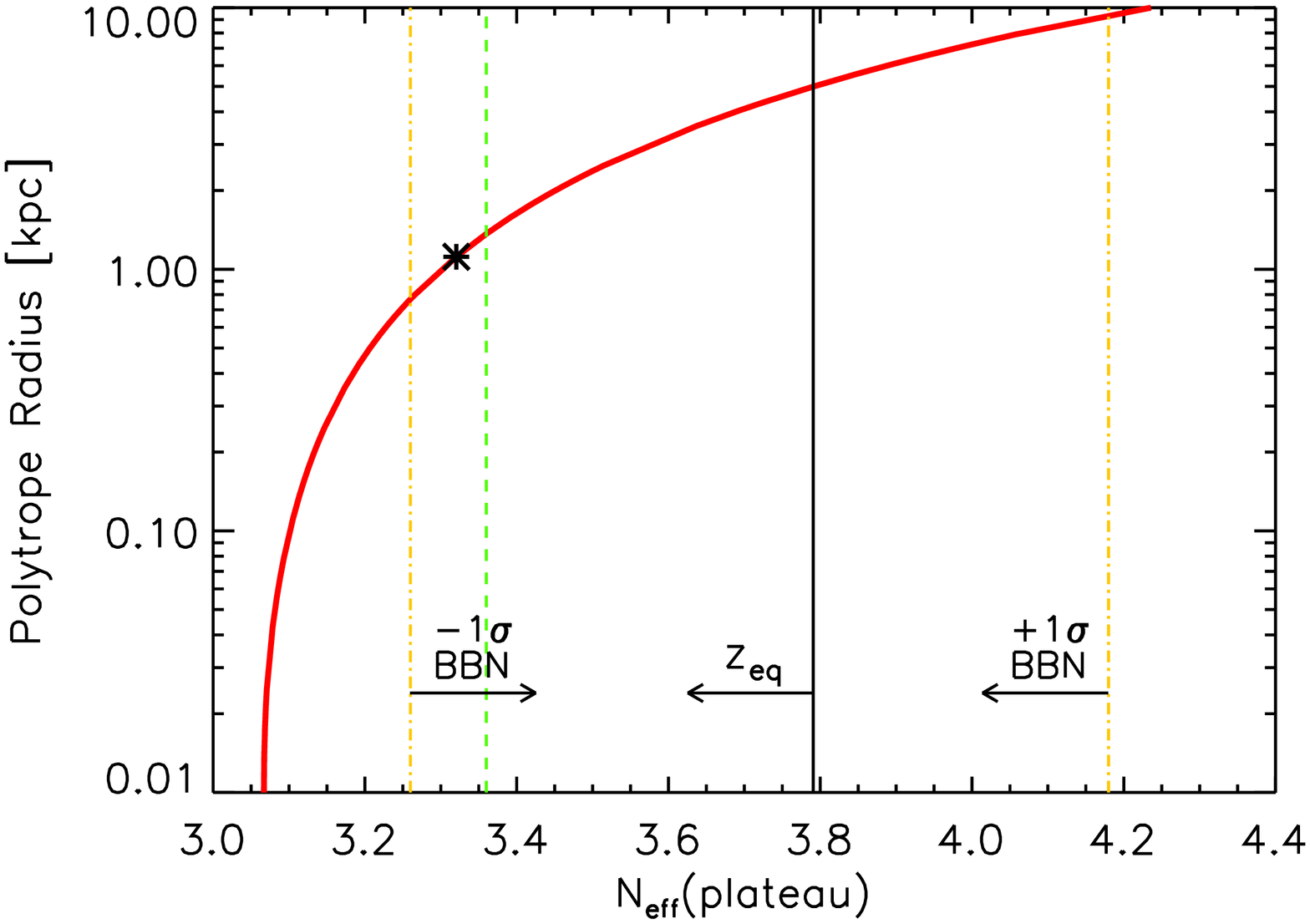}

    \end{minipage}%
 \caption{(Color online) 
Radius of a virialized, polytropic SFDM halo, which forms during the matter-dominated era, 
as a function of $N_{\rm{eff}}$ during the radiation-dominated era. The relation is shown by the solid curve, 
on which the star represents our fiducial model. 
The polytrope radius is considered as the minimum length scale of structures. 
The two dash-dotted vertical lines indicate the $1\sigma$ limits of $N_{\rm{eff}}$ from BBN measurements, 
while the dashed vertical line indicates the central value of $N_{\rm{eff}}$ from CMB measurements 
(the latter is the same as in Fig. \ref{figure:Comparison}, lower plots). 
The solid vertical line denotes the upper bound of $N_{\rm{eff}}$ during the plateau 
so as to fulfill the constraint from fixed $z_{\rm{eq}}$. 
 }
 \label{figure:NeffRadius}
\end{figure*}

On the other hand, we have verified in Section \ref{sec:FRWbackground4} that 
as $N_{\rm{eff}}$ reaches the plateau (SFDM reaches the radiationlike phase), 
its value is also purely determined by $\lambda/(mc^2)^2$. Therefore, we can plot 
the polytrope radius against $N_{\rm{eff}}$ corresponding to the plateau, revealing a hitherto unnoticed 
relation between the scale of the smallest dark matter structures and the number of relativistic species
in the radiation-dominated era, see Figure \ref{figure:NeffRadius}.

The plot shows that higher $N_{\rm{eff}}$ implies stronger self-interaction pressure hence larger 
minimum scale for dark matter structure. The constraints discussed in the above section gives 
the allowed window of the minimum length scale, which is the segment of the curve between 
the left dotted vertical, the lower 1$\sigma$ limit from BBN measurement, and the solid vertical, 
the bound from the constraint on $\lambda/(mc^2)^2$ by $z_{\rm{eq}}$, 
see equation (\ref{equation:constraint1}). We can see that our fiducial model which corresponds 
to a minimum length scale of 1.1 kpc lies within the allowed window. 
It is a satisfactory result since this is about the scales where the small-scale CDM 
problems start to be significant from observations \cite{dB2001, Oh2008, Amorisco2012}. 
We should also note that the allowed window for the minimum length scale 
is subject to changes in future observational results from CMB and BBN.

\subsection{Imprints on the CMB from a time varying $N_{\rm{eff}}$}
\label{sec:neffcmb}

Besides BBN, the angular power spectrum of the CMB temperature
fluctuations can also be used to constrain the expansion rate of the
universe during the radiation-dominated era by the ratio of the Silk damping scale $\theta_D$ 
to the sound horizon scale $\theta_*$ \cite{Planck}. 
This provides a different constraint on $N_{\rm{eff}}$ from that described above from BBN. 
While the expansion rate depends upon the number of relativistic species present as well, 
it should be noted, though, that because of its possible evolution, 
the $N_{\rm{eff}}$ affecting the CMB power spectrum 
is not the same as the $N_{\rm{eff}}$ during BBN. 
The former concerns its value during the epoch spanned by the moment at which the 
smallest angular scale probed ($l\sim3000$) enters the horizon, 
$z_{\rm{entry}}\sim6\times10^4$, to that of matter-radiation equality at $z_{\rm{eq}}\sim3\times10^3$, 
as pointed out in Ref. \cite{Bashinsky2004}. By contrast, the BBN constraint 
probes $N_{\rm{eff}}$ at $z\gtrsim z_{\rm{nuc}}\sim3\times10^8$. 
In the $\Lambda$SFDM model, $N_{\rm{eff}}$ evolves over time in such a way that 
$N_{\rm{eff}}$ is (at most) its plateau value at $z_{\rm{entry}}$, 
and finally reduces to the standard value of 3.046 when SFDM becomes
fully non-relativistic (before $z_{\rm{eq}}$, as explained in Section
\ref{sec:FRWbackground3}). Therefore, the plateau value of 
$N_{\rm{eff}}$ during the radiation-dominated era serves as an upper bound for what is 
responsible for the expansion rate from $z_{\rm{entry}}$ to $z_{\rm{eq}}$. 

However, a complication arises that the ratio of $\theta_D/\theta_*$  
does not only depend on the expansion rate during the period mentioned above, 
but also on the primordial Helium abundance $Y_P$, since the damping tail
is subject to the number density of free electrons $n_e$ \cite{WMAP9}. 
Actually, 
\begin{equation}\label{equation:scaleratio}
	\theta_D/\theta_*\propto\sqrt{\frac{H}{n_e}}\propto\sqrt{\frac{H}{1-Y_P}},
\end{equation}
where $H$ refers to the Hubble expansion rate between $z_{\rm{entry}}$ and $z_{\rm{eq}}$. 
We also know that $Y_P$ is dependent upon $N_{\rm{eff}}$ during BBN 
(an increase in $N_{\rm{eff}}$ results in a higher $Y_P$).  
Therefore, the relativistic degrees of freedom suggested by CMB measurements, 
e.g., $N_{\rm{eff}}=3.36\pm0.34$ given by Planck+WP+highL \cite{Planck}, 
(again, models with \textit{constant} $N_{\rm{eff}}$ are fitted to the data) 
is in fact an imprint from both
$N_{\rm{eff}}$ during BBN at early times (through $Y_P$) and its later evolution 
from $z_{\rm{entry}}$ to $z_{\rm{eq}}$ (through $H$), in $\Lambda$SFDM. 
In fact, equation (\ref{equation:scaleratio}) implies that $\theta_D/\theta_*$ increases 
when either $H$ or $Y_P$ increases, provided a higher $N_{\rm{eff}}$ at the respective epoch, 
which then suggests that, $N_{\rm{eff}}$ given by CMB measurements be between 
the $N_{\rm{eff}}$ during BBN and the $N_{\rm{eff}}$ from $z_{\rm{entry}}$ to $z_{\rm{eq}}$ 
(the exact relation requires the calculation of linear growth). 
We then note that SFDM naturally provides an explanation for 
the difference between the $N_{\rm{eff}}$ values currently measured from
BBN and CMB, in which the BBN value is larger than the CMB value.

\subsection{Early stiff-matter phase}
\label{sec:FRWbackground5}

We have seen in Section \ref{sec:SFDMbackground2} that SFDM undergoes a stiff phase, when $\bar p_{\rm{SFDM}} \approx \bar\rho_{\rm{SFDM}}$ and $\bar \rho_{\rm{SFDM}} \propto a^{-6}$.  
This feature of scalar fields has been noted before in models where the scalar field describes post-inflation universe or dark energy; see, e.g., Refs. \cite{RP1988, Joyce1997, CEM2007}. 
In Ref. \cite{ALS2002}, this feature has been found for SFDM without self-interaction. However, these authors did not find the accompanying constraint on the particle parameters, and also were limited to analytic treatment, while we calculated the evolution numerically and explore the parameter space where the stiff phase is important. The first suggestion of a stiff equation of state for the baryonic fluid in the early universe seems to be by Refs. \cite{Zeldovich1962, Zeldovich1972}. The possibility of pre-BBN non-standard expansion histories, which includes a component decaying as $a^{-6}$, has been considered, e.g. in Refs. \cite{KT1990} and \cite{Dutta2010}. However, the stiff components studied there do not undergo any transition, i.e., always decay as $a^{-6}$, unlike our model.     

In a $\Lambda$SFDM universe, the stiff phase can last until BBN occurs 
due to the constraints on the expansion rate. As we have seen in Section \ref{sec:FRWbackground4}, 
for all viable models the stiff phase completely ends before $a_{\rm{nuc}}$.
An interesting question is whether the stiff phase before $a_{\rm{n/p}}$ 
will affect baryonic processes so as to leave an imprint on BBN products. 
In fact, the free neutron abundance is subject to beta decay, 
which happens ever since neutrons have existed, going as
$e^{-t/\tau_n}$ with the neutron decay time $\tau_n$. Thus, the
number of free neutrons left for nucleosynthesis depends on the age of the Universe, $t$,
since the QCD phase transition. Now, if $t = 1/(3H)$ in the stiff phase, instead of
the radiation-era dependence, $t = 1/(2H)$, this will change the
number of available free neutrons before $a_{\rm{nuc}}$. The change
in the age of the Universe is marginal, though, with a factor of $1/3$, instead of $1/2$,
to multiply the decay-factor. As shown in the left-hand plot of Figure \ref{figure:Hubble}, the Hubble time
at the epoch of the stiff-radiation transition is $\lesssim 1s \ll \tau_n$, which actually applies to all viable models. 
It is thus safe enough to constrain SFDM parameters only during BBN.

As far as the QCD phase transition is concerned, which happens somewhere
between 150 -- 300 MeV, there is still a lot of ongoing work to
understand those processes in full. However, the relaxation time of the
strong force is so tiny, in contrast to the Hubble time, that the QCD transition takes
place in chemical equilibrium all the time, without a freeze-out timing issue. 
Therefore, we think the universe can be in the SFDM-dominated era in the stiff phase 
with a higher expansion rate, as suggested by our model, and yet
accomplish a standard hadron era.

\subsection{Implications for fuzzy dark matter} \label{sec:fuzzy}

Our analysis above is valid for arbitrary value of $\lambda$. It is natural, therefore, 
to ask what the implications of our constraints are for the limiting case of $\lambda=0$. 
SFDM without self-interaction, $\lambda \equiv 0$, or fuzzy dark matter, 
is left with the quadratic potential in (\ref{equation:potential}). 
Its popularity is reflected by numerous previous investigations; see, e.g., Refs. \cite{KMZ, Sin, Magana2012}. 
One reason is that, even without the self-interaction pressure associated with nonzero $\lambda$, 
FDM still provides a mechanism to suppress structures on scales below 
$\lambda_{\rm{deB}}$, as a result of quantum pressure due to 
the Heisenberg Uncertainty Principle. Since it is an important
special case of the model we have investigated, we devote
this subsection to summarize the implications for this model from
our analysis.

Without the self-interaction term, FDM has only two evolutionary phases,
the early relativistic, stiff-matter phase, followed by the non-relativistic, CDM-like phase. 
For values of $m$ which are large enough to make this transition occur before 
the BBN epoch, the redshift of matter-radiation equality, $z_{\rm{eq}}$, is unaffected
because of the absence of the plateau (the radiationlike phase). 
Nevertheless, BBN sets a constraint on the only parameter left, the mass $m$.
Since the kinetic term in the SFDM energy density (\ref{equation:densitybu}) 
goes inversely with $m$, 
the transition between stiff and dust-like equation of state happens \textit{later}
with \textit{decreasing} mass. In fact, according to Figure \ref{figure:PPS}, if we accept 
the 1$\sigma$ limits on $N_{\rm{eff}}$ allowed by BBN to constrain $m$, 
there is \textit{no} value of $m$ for which $\lambda=0$ can be consistent, which indicates 
a rejection of the FDM model at $\ge1\sigma$. 
We highlight this result, since FDM with $m \sim 10^{-23}-10^{-22}$ eV/$c^2$ has
been a very popular candidate in the literature because the minimum length scale 
$\sim\lambda_{\rm{deB}}$ that corresponds to such particle mass is roughly 1 kpc, 
as mentioned in the introduction. Again, we should admit that, 
placing a less tight constraint, e.g., within 2$\sigma$, 
FDM may be able to fit BBN measurements.

\section{Conclusions}  \label{sec:conclusion}

We presented the cosmological evolution of a universe in which dark matter 
is comprised of ultralight self-interacting bosonic particles, which form a
Bose-Einstein condensate, described by a classical complex scalar field
(SFDM). We solved the Klein-Gordon and Einstein field equations for
the time-dependence of an FRW universe with this form of dark
matter, and placed constraints on the SFDM particle mass $m$ and
self-interaction coupling strength $\lambda$ (or equivalently $\lambda/(mc^2)^2$) 
from cosmological observations.

Unlike standard CDM, which is always non-relativistic
once it decouples from the background, SFDM has an evolving equation of state. 
As a result, there are four eras in the evolution of a homogeneous $\Lambda$SFDM universe: 
the familiar radiation-dominated, matter-dominated and Lambda-dominated
eras common to standard $\Lambda$CDM as well, but also an earlier era 
dominated by SFDM with a stiff equation of state. 
Then, $\bar{p} \simeq \bar{\rho} \propto a^{-6}$ and $a \propto t^{1/3}$. 
The manifestation of this era does not depend on 
whether self-interaction has been included or not. 
It appears in fuzzy dark matter models with $\lambda \equiv 0$ as well. 
The timing and longevity of this era (or the stiff phase of SFDM), however, 
depend on the particular values of SFDM particle parameters, 
$m$ along with $\lambda/(mc^2)^2$. It is necessary to ensure that 
the stiff phase is ending when big bang nucleosynthesis begins. 
This finding is a special novelty of our analysis. At intermediate times, SFDM is radiationlike, 
with $\bar{p} \simeq \bar{\rho}/3$. Finally, SFDM must transition to the CDM-like phase 
before the epoch of matter-radiation equality, and thereafter behaves as a pressureless dust.

The effect of this SFDM equation of state evolution on the expansion rate and
mass-energy content of the universe enables us to place constraints
on $m$ and $\lambda/(mc^2)^2$, by using $N_{\rm{eff}}$ at BBN, and $z_{\rm{eq}}$ 
measured by CMB anisotropy. We find that $m\ge2.4\times10^{-21}$eV$/c^2$ and 
$9.5\times10^{-19}\rm{eV^{-1}cm^3}\le\lambda/(mc^2)^2\le4\times10^{-17}\rm{eV^{-1}cm^3}$. 
While we are able to place more stringent bounds 
on these particle parameters than the previous literature, there remains 
a large range of SFDM parameters which provides an expansion history 
in conformity with cosmological observations. Our investigations thereby 
contribute to previous efforts in establishing SFDM as a viable dark matter candidate. 
Work is in progress to study the linear and nonlinear
growth of structures in a $\Lambda$SFDM universe, in order
to find out which part of the parameter space of SFDM is able to
explain observations on all scales self-consistently.

\begin{acknowledgments}
We are grateful for helpful discussions with Can Kilic, Dustin
Lorshbough, Tonatiuh Matos and Eiichiro Komatsu. This work was
supported in part by NSF grant AST-1009799 and NASA grant NNX11AE09G
to PRS. TRD also acknowledges support by the Texas Cosmology Center
of the University of Texas at Austin.
\end{acknowledgments}

\appendix

\section{Basic equations in a perturbed Friedmann-Robertson-Walker
metric}\label{app:pfrw}

The general perturbed Friedmann-Robertson-Walker (FRW) metric in the
comoving frame has the form
\begin{displaymath}
    \ud s^2=(1+2\Psi/c^2)c^2\ud t^2-2a(t)w_ic\ud t\ud x^i-
\end{displaymath}
\begin{equation}
 - a^2(t)[(1-2\Phi/c^2)\delta_{ij}+H_{ij}]\ud x^i\ud x^j,
\end{equation}
where the perturbed quantities $|\Psi|/c^2,|\Phi|/c^2,|w_i|$, and
$|H_{ij}|$ are all $\ll 1$.

\subsection{Conformal Newtonian gauge} \label{app:newtoniangauge}
We can apply the conformal Newtonian Gauge if only scalar
perturbations are permitted, where the metric reduces to \cite{MBW}
\begin{equation} \label{equation:metric}
    \ud s^2=(1+2\Psi/c^2)c^2\ud t^2-a^2(t)(1-2\Phi/c^2)\delta_{ij}\ud x^i\ud x^j, 
\end{equation}
or
\begin{displaymath}
    g_{00} =  1+2\frac{\Psi}{c^2},~ g_{ij} =  -a^2(t)\left(1-2\frac{\Phi}{c^2}\right)\delta_{ij},
\end{displaymath}
\begin{displaymath}
    g^{00} =  1-2\frac{\Psi}{c^2},~ g^{ij} = -\frac{1}{a^2(t)}\left(1+2\frac{\Phi}{c^2}\right)\delta_{ij}
\end{displaymath}
The corresponding Christoffel symbols are \cite{Dodelson}
\begin{displaymath}
    \Gamma^0_{~00} =  \frac{1}{c^3}\partial_t\Psi,~\Gamma^0_{~i0} =  \frac{1}{c^2}\partial_i\Psi,~\Gamma^i_{~00} = \frac{1}{c^2a^2}\partial_i\Psi,
\end{displaymath}
\begin{displaymath}
    \Gamma^i_{~j0}  =  \left(-\frac{1}{c^3}\partial_t\Phi+\frac{\ud a/\ud t}{ca}\right)\delta_{ij},
\end{displaymath}
\begin{displaymath}
\Gamma^0_{~jk}  = \left(-\frac{a^2}{c^3}\partial_t\Phi+\frac{a\ud
a/\ud
t}{c}(1-2\frac{\Psi}{c^2}-2\frac{\Phi}{c^2})\right)\delta_{jk},
\end{displaymath}
\begin{equation} \label{equation:cf}
    \Gamma^i_{~jk}=-\frac{1}{c^2}\left(\partial_k\Phi\delta_{ij}+\partial_j\Phi\delta_{ik}-\partial_i\Phi\delta_{jk}\right).
\end{equation}

\subsection{Klein-Gordon equation} \label{app:kgeq}

The variation of the action
\begin{equation}
    \mathscr{S}=\int \ud^4x\sqrt{-g}\mathscr{L}(\psi, \psi^*, \partial_\mu\psi, \partial_\mu\psi^*),
\end{equation}
with $g=\text{det}(g_{\mu\nu})$, yields
\begin{displaymath}
    \delta\mathscr{S} = \int \ud^4x\sqrt{-g}~\times
\end{displaymath}
\begin{displaymath}
  \times \left(\frac{\partial\mathscr{L}}{\partial(\partial_\mu\psi)}\partial_\mu\delta\psi+
    \frac{\partial\mathscr{L}}{\partial\psi}\delta\psi+\frac{\partial\mathscr{L}}{\partial(\partial_\mu\psi^*)}\partial_\mu\delta\psi^*
    +\frac{\partial\mathscr{L}}{\partial\psi^*}\delta\psi^*\right)
\end{displaymath}
\begin{displaymath}
     =  \int \ud^4x\bigg[\left(-\partial_\mu\left(\sqrt{-g}\frac{\partial\mathscr{L}}{\partial(\partial_\mu\psi)}\right)+
     \sqrt{-g}\frac{\partial\mathscr{L}}{\partial\psi}\right)\delta\psi+
\end{displaymath}
\begin{equation}\label{equation:rCl}
+\left(-\partial_\mu\left(\sqrt{-g}\frac{\partial\mathscr{L}}{\partial(\partial_\mu\psi^*)}\right)+
\sqrt{-g}\frac{\partial\mathscr{L}}{\partial\psi^*}\right)\delta\psi^*\bigg].
\end{equation}
For arbitrary $\delta\psi$ and $\delta\psi^*$, $\delta\mathscr{S} =
0$ only when both integrands in the expression above are constantly
zero, which yields the Euler-Lagrangian equation
\begin{equation}
    \frac{1}{\sqrt{-g}}\partial_\mu\left(\sqrt{-g}\frac{\partial\mathscr{L}}{\partial(\partial_\mu\psi)}\right)-\frac{\partial\mathscr{L}}{\partial\psi}=0.
\end{equation}
Upon inserting the Lagrangian (\ref{equation:lag}), one recovers
(\ref{equation:relkg}).

\subsection{Einstein field equations and curvature tensor}
\label{app:efe}

The Einstein-Hilbert action is defined as
\begin{equation}
    \mathscr{S}_H=\int \ud^4x\sqrt{-g}\left(\frac{R}{16\pi Gc^{-4}}+\mathscr{L}\right).
\end{equation}
The Einstein field equations can be derived from the principle of
least action with variation in $g^{\mu\nu}$:
\begin{IEEEeqnarray}{rCl}
    0 & = & \delta\mathscr{S}_H\\
    & = & \int d^4x\left(\frac{\delta(\sqrt{-g}R)}{16\pi Gc^{-4}}+\delta(\sqrt{-g}\mathscr{L})\right)\nonumber\\
    & = & \int d^4x\bigg(-\frac{\sqrt{-g}}{16\pi Gc^{-4}}(R_{\mu\nu}-\frac{1}{2}g_{\mu\nu}R)+\nonumber\\
    & & +\frac{\delta(\sqrt{-g}\mathscr{L})}{\delta g^{\mu\nu}}\bigg)\delta g^{\mu\nu}.\nonumber
\end{IEEEeqnarray}
Defining the energy-momentum tensor as
\begin{equation} \label{stresstensor}
    T_{\mu\nu}\equiv
    \frac{2}{\sqrt{-g}}\frac{\delta\left(\sqrt{-g}\mathscr{L}(g^{\mu\nu},\partial^\rho g^{\mu\nu})\right)}{\delta g^{\mu\nu}}=
    2\frac{\delta\mathscr{L}}{\delta
    g^{\mu\nu}}-g_{\mu\nu}\mathscr{L},
\end{equation}
the field equations are thus
\begin{equation}
    R_{\mu\nu}-\frac{1}{2}g_{\mu\nu}R=\frac{8\pi G}{c^4}T_{\mu\nu}.
\end{equation}
The Riemann curvature tensor is defined as
\begin{equation}
    R^\rho_{~\sigma\mu\nu}=\partial_\mu\Gamma^\rho_{~\sigma\nu}-\partial_\nu\Gamma^\rho_{~\sigma\mu}+\Gamma^\rho_{~\mu\alpha}\Gamma^\alpha_{~\sigma\nu}-\Gamma^\rho_{~\nu\alpha}\Gamma^\alpha_{~\sigma\mu}.
\end{equation}
With the Christoffel symbols (\ref{equation:cf}) we can calculate
the diagonal Ricci tensors to first order in
$|\Psi|/c^2,~|\Phi|/c^2$,
\begin{displaymath}
    R_{\mu\nu} \equiv  R^\rho_{~\mu\rho\nu},
\end{displaymath}
\begin{displaymath}
    R_{00}  =  -\frac{3}{c^2}\frac{\ud^2a/\ud t^2}{a}+\frac{1}{c^2a^2}\nabla^2\Psi+\frac{3}{c^4}\partial^2_t\Phi+
\end{displaymath}
\begin{displaymath}
    + \frac{3\ud a/\ud t}{c^4a}(\partial_t\Psi+2\partial_t\Phi),
\end{displaymath}
\begin{displaymath}
    R_{ii}  =  \frac{a\ud^2a/\ud t^2+2(\ud a/\ud t)^2}{c^2}\left(1-2\frac{\Psi}{c^2}-2\frac{\Phi}{c^2}\right)-
\end{displaymath}
\begin{displaymath}
 - \frac{a\ud a/\ud t}{c^4}(6\partial_t\Phi+\partial_t\Psi)-\frac{a^2}{c^4}\partial^2_t\Phi+\frac{1}{c^2}\nabla^2\Phi-\frac{1}{c^2}\partial^2_i(\Psi-\Phi).\nonumber
\end{displaymath}
Consequently the Ricci scalar is
\begin{IEEEeqnarray}{rCl}
    R & \equiv & g^{\mu\nu}R_{\mu\nu}= -\frac{6}{c^2}\left(\frac{\ud^2a/\ud t^2}{a}+\frac{(\ud a/\ud t)^2}{a^2}\right)+\nonumber\\
    & & +\frac{2}{c^2a^2}\nabla^2(\Psi-\Phi)- \frac{2}{c^2a^2}\nabla^2\Phi+\frac{6\partial^2_t\Phi}{c^4}+\nonumber\\
    & & +\frac{6\ud a/\ud t}{c^4a}(\partial_t\Psi+4\partial_t\Phi)+\nonumber\\
    & & + \frac{12}{c^4}\left(\frac{\ud^2a/\ud t^2}{a}+\frac{(\ud a/\ud t)^2}{a^2}\right)\Psi.\nonumber
\end{IEEEeqnarray}

\section{Oscillation and charge of the complex scalar field in an homogeneous Friedmann-Robertson-Walker metric}
\label{app:charge}

Let us write the equation of motion with homogeneous FRW metric (\ref{equation:febu}), again, 
\begin{equation}
    \frac{\hbar^2}{2mc^2}\partial^2_t\psi+\frac{\hbar^2}{2mc^2}\frac{3\ud a/\ud t}{a}\partial_t\psi+\frac{1}{2}mc^2\psi+\lambda|\psi|^2\psi=0.
\end{equation}
Now, we will decompose the complex scalar field as
\begin{equation}
	\psi=|\psi|e^{i\theta},
\end{equation}
where $|\psi|$ is the amplitude of the scalar field and $\theta$ is its phase. They are both real functions. 
Inserting this decomposition into the equation of motion above yields
\begin{IEEEeqnarray}{l}
	\frac{\hbar^2}{2mc^2}\left(\partial^2_t|\psi|-|\psi|(\partial_t\theta)^2\right)+\frac{\hbar^2}{2mc^2}\frac{3\ud a/\ud t}{a}\partial_t|\psi|+\nonumber\\
	+\frac{1}{2}mc^2|\psi|+\lambda|\psi|^3=0,\label{equation:deom1}\\
	\frac{\hbar^2}{2mc^2}\left(2\partial_t|\psi|\partial_t\theta+|\psi|\partial^2_t\theta\right)+\nonumber\\
	+\frac{\hbar^2}{2mc^2}\frac{3\ud a/\ud t}{a}|\psi|\partial_t\theta=0.\label{equation:deom2}
\end{IEEEeqnarray}

We first look at equation (\ref{equation:deom1}). It is the phase that carries the major oscillation behavior 
for a complex scalar field, while the time dependence of the amplitude is smooth. 
In the fast oscillation regime, in which the Hubble expansion rate 
$H=\frac{\ud a/\ud t}{a}$ is minute compared with $\partial_t\theta$, 
we also assume that $\partial_t|\psi|/|\psi|\ll\partial_t\theta$ (which is not always the case). 
We can then neglect the terms involving $\partial_t|\psi|$ and $H$ in equation (\ref{equation:deom1}) and obtain 
\begin{equation}
	-\frac{\hbar^2}{2mc^2}|\psi|(\partial_t\theta)^2+\frac{1}{2}mc^2|\psi|+\lambda|\psi|^3=0.
\end{equation}
We define the angular oscillation frequency as $\omega\equiv\partial_t\theta$. 
Rearranging the equation above yields
\begin{equation} \label{equation: oscillationfrequency}
	\omega=\frac{mc^2}{\hbar}\sqrt{1+\frac{2\lambda}{mc^2}|\psi|^2}, 
\end{equation}
which can be also viewed as the dispersion relation in the zero-momentum case of our complex scalar field. 
We should bear in mind that the relation (\ref{equation: oscillationfrequency}) 
is only valid when $\omega\gg H$. In the case of a free field 
($\lambda=0$), the frequency reduces to the particle mass, $\omega=mc^2/\hbar$, as one may expect.

Let us turn to equation (\ref{equation:deom2}). It can be exactly integrated once \cite{GuHwang2001}, giving 
\begin{IEEEeqnarray}{rl}
	& \partial_t(a^3|\psi|^2\partial_t\theta)=0.\nonumber
\end{IEEEeqnarray}
Therefore, we can see that $a^3|\psi|^2\partial_t\theta$ is conserved over cosmic time. 
In fact, it is proportional to the conserved charge density $Q$, defined in Section \ref{sec:BEC}, 
\begin{equation}
	a^3|\psi|^2\partial_t\theta=Q\frac{mc^2}{\hbar}.
\end{equation}
In the case of a complete BEC, anti-bosons are nearly annihilated away so that the charge 
basically equals the total number of condensed bosons (see Refs. \cite{Kapusta1981, MMPP2001, Briscese2011}). 
The conservation of the charge, or equivalently, the conservation of the charge density $Q$ 
results from the global U(1) symmetry of the Lagrangian density (\ref{equation:lag}).
This is a distinct feature in contrast to a real scalar field. Since a real field does not possess phase information 
$\theta$, there is no global U(1) symmetry and no non-trivial charge. 
In fact, the boson is its own anti-boson for a real scalar field.

\section{Matching conditions of the early-time and late-time solution}\label{app:matching}

The integration of the equations for the early-time solution is performed backwards in time from the matching point with the late-time solution, at $\omega/H=200$. This matching condition requires that the starting values of $\bar p$, $\bar\rho$ and the scale factor $a$ for the early-time solution are given by $\langle\bar p\rangle$, $\langle\bar\rho\rangle$ and $a$ at the matching point in the late-time solution, with the value of $B$ there set as follows (we omit 
the subscript SFDM in this appendix). The starting value of $B$ should be determined, in principle, by the following equation. 
Summing equations (\ref{equation:densitybu}) and (\ref{equation:pressurebu}) yields 
\begin{IEEEeqnarray}{rCl}\label{equation:matchB}
	\bar\rho+\bar p & = & \frac{\hbar^2}{mc^2}|\partial_t\psi|^2\nonumber\\
	& = & \frac{\hbar^2}{mc^2}\left((\partial_t|\psi|)^2+|\psi|^2(\partial_t\theta)^2\right)\nonumber\\
	& = & \frac{\hbar^2}{mc^2}\left(\frac{(\partial_t|\psi|^2)^2}{4|\psi|^2}+\frac{(|\psi|^2\partial_t\theta)^2}{|\psi|^2}\right)\nonumber\\
	& = & \frac{\hbar^2}{mc^2|\psi|^2}\left(\frac{1}{4}\left(\frac{B}{mc^2}\right)^2+\frac{(Qmc^2/\hbar)^2}{a^6}\right)\nonumber\\
	& = & \frac{\hbar^2}{2(\bar\rho-\bar p)}\left(\sqrt{1+\frac{4\lambda(\bar\rho-\bar p)}{(mc^2)^2}}+1\right)\times\nonumber\\
	& & \times\left(\frac{1}{4}\left(\frac{B}{mc^2}\right)^2+\frac{(Qmc^2/\hbar)^2}{a^6}\right).
\end{IEEEeqnarray}
Therefore, if we know the conserved charge density $Q$ precisely, we should be able to 
calculate the exact value of $B$. Unfortunately, this is not practical, 
for $Q$ is so huge (for a BEC) that the last term on the right-hand side of equation (\ref{equation:matchB}) 
is greater than the term involving $B$ by many orders of magnitude. 
As a matter of fact, in the fast oscillation regime, the term involving $B$ is always subdominant to the term involving $Q$ for a BEC, justifying our assumption that $\partial_t|\psi|/|\psi|\ll\partial_t\theta$ (in the slow oscillation regime it is the converse). 
Thus, though we know that $Q\approx\bar\rho_{\rm{dm}}(t_0)/(mc^2)$, 
even a tiny error in $Q$ will lead to a big variation in the value of $B$, 
making it impossible to use equation (\ref{equation:matchB}) to determine $B$. 

Nevertheless, we have confirmed by testing different starting values of $B$ that, changing $B$ does not affect the time-average values of the SFDM energy density $\bar\rho$ and pressure $\bar p$, only causing different oscillation amplitudes of $\bar p$. Recall that the evolution of $\bar\rho$ is always smooth (see Section \ref{sec:method}). The expansion history of the homogeneous background universe is thus unaffected despite the uncertainty in $B$, since the Friedmann equation (\ref{equation:friedmannbu}) only concerns $\bar\rho$, and hence we are free to choose the starting value of $B$, within the range derived from equation (\ref{equation:matchB}), which corresponds to the range of uncertainty in the exact value of $Q$. 
Here is another remarkable feature of the complex scalar field: although the SFDM pressure shows oscillation generically, the amplitude of this oscillation is always a small fraction of the mean value for subdominant $B$ values, as oscillations mainly manifest in the phase. This is distinct from the real field case again, as for a real scalar field, $\bar w=\bar p/\bar\rho$ oscillates between $-1$ and $1$ (see Ref. \cite{Magana2012}). 

In this work, we choose the starting value of $B$ for the early-time solution in a way that makes the early-time solution smoothly match onto the late-time solution, particularly in $\bar p$, with zero oscillation amplitude. To see that, subtracting equation (\ref{equation:pressurebu}) from equation (\ref{equation:densitybu}) yields 
\begin{IEEEeqnarray}{rCl}
	B & = & mc^2\partial_t|\psi|^2=\frac{mc^2\partial_t(\bar\rho-\bar p)}{mc^2+2\lambda|\psi|^2}\nonumber\\
	& = & \frac{\partial_t(\bar\rho-\bar p)}{\sqrt{1+4\lambda(\bar\rho-\bar p)/(mc^2)^2}}. 
\end{IEEEeqnarray}
The starting value of $B$ is then taken as 
\begin{IEEEeqnarray}{rCl}
	B_{\rm{match}} & = & \frac{\partial_t(\langle\bar\rho\rangle-\langle\bar p\rangle)}{\sqrt{1+4\lambda(\langle\bar\rho\rangle-\langle\bar p\rangle)/(mc^2)^2}}\nonumber\\
	& = & \frac{\partial_t\langle\bar\rho\rangle(1-\partial\langle\bar p\rangle/\partial\langle\bar\rho\rangle)}{\sqrt{1+4\lambda(\langle\bar\rho\rangle-\langle\bar p\rangle)/(mc^2)^2}}\nonumber\\
	& = & -\frac{\ud a/\ud t}{a}\frac{(\langle\bar\rho\rangle+\langle\bar p\rangle)}{\sqrt{1+4\lambda(\langle\bar\rho\rangle-\langle\bar p\rangle)/(mc^2)^2}}\times\nonumber\\
	& & \times\left(2+\frac{1}{\sqrt{1+6\lambda\langle\bar\rho\rangle/(mc^2)^2}}\right) 
\end{IEEEeqnarray}
where we assume that the matching point lies within the radiationlike phase of SFDM. With such starting value of $B$, the derived evolution of $\bar w=\bar p/\bar\rho$ from the integration of the exact equations connects smoothly to that of the late-time solution for $\langle\bar p\rangle/\langle\bar\rho\rangle$, with no oscillation, as shown in the right-hand plot of Figure \ref{figure:EOSplots}.

\bibliography{SFDM_cosmology}

\end{document}